\documentclass[journal,twoside]{IEEEtran}

\usepackage{amsmath,amssymb,amsfonts}
\usepackage{algorithmic}
\usepackage{algorithm}
\usepackage{array}
\usepackage{stfloats}
\usepackage{url}
\usepackage{verbatim}
\usepackage{textcomp}
\usepackage{graphicx}
\usepackage{cite}
\usepackage{multirow}
\usepackage{xcolor}
\usepackage{xspace}
\usepackage{hyperref}
\hypersetup{hidelinks}

\usepackage[english]{babel}
\usepackage{threeparttable}
\usepackage{float}
\usepackage{flafter}
\usepackage{comment}
\usepackage{tabularx}
\usepackage{booktabs}
\usepackage[utf8]{inputenc}
\usepackage{makecell}     
\usepackage{pifont}       
\usepackage{wasysym}      
\usepackage{tikz}
\usepackage[defaultlines=1,all]{nowidow}

\usepackage[skip=12pt]{caption}
\usepackage{subcaption}
\graphicspath{{./pdf/}{./jpeg/}{./fig/}{./visio/}{./pics/}}
\DeclareGraphicsExtensions{.pdf,.jpeg,.jpg,.png}

\newcommand{\cmark}{\ding{51}}%
\newcommand{\xmark}{\ding{55}}%

\newcommand{\com}[1]{} 

\newcommand{\head}[1]{{\noindent \bf #1}}
\newcolumntype{C}[1]{>{\centering\arraybackslash}m{#1}}

\begin{document}

\widowpenalty=10
\clubpenalty=10
\brokenpenalty=10

\setlength{\belowdisplayskip}{0.5mm}
\setlength{\belowdisplayshortskip}{0mm}
\setlength{\abovedisplayskip}{0.5mm}
\setlength{\abovedisplayshortskip}{0mm}
\setlength{\jot}{0.2mm}
\thickmuskip=0.5\thickmuskip

\setlength{\tabcolsep}{3pt}

\def\BibTeX{{\rm B\kern-.05em{\sc i\kern-.025em b}\kern-.08em
    T\kern-.1667em\lower.7ex\hbox{E}\kern-.125emX}}
\markboth{}{Zhu \MakeLowercase{\textit{et al.}}: Scaling WiFi Sensing for Ubiquitous Home Monitoring}

\title{Scaling WiFi Sensing for Ubiquitous Home Monitoring: Lessons from Real-World Deployment on Millions of Devices}

\author{Guozhen Zhu,~\IEEEmembership{Member,~IEEE,}\thanks{ Manuscript submitted May,~2026.  \textit{(Corresponding author: Guozhen Zhu.)} } Yuqian Hu,~\IEEEmembership{Member,~IEEE,} Chenshu Wu,~\IEEEmembership{Senior Member,~IEEE,},  Wei-Hsiang Wang,~\IEEEmembership{Member,~IEEE,}
 Beibei Wang,~\IEEEmembership{Fellow,~IEEE,}  K. J. Ray Liu,~\IEEEmembership{Fellow,~IEEE}
\thanks{
Guozhen Zhu, Beibei Wang, Wei-Hsiang Wang, Yuqian Hu and K. J. Ray Liu are with ADT Research, Rockville, MD, 20852 USA. Chenshu Wu is with the Department of Computer Science, The University of Hong Kong, Hong Kong CN. (e-mail: gzzhu@terpmail.umd.edu;  \{yuqian.hu, beibei.wang, ray.liu\}@originwirelessai.com; chenshu@cs.hku.hk). }}

\maketitle

\begin{abstract}
WiFi-based home monitoring offers compelling advantages over traditional camera and sensor solutions by leveraging existing wireless infrastructure for contactless, privacy-preserving and through-the-wall detection. This paper presents insights from developing and deploying a WiFi-based human monitoring system across real-world residential environments, addressing the gap between academic research and practical deployment. Through a two-year study involving 280 edge devices across 15 homes in 11 U.S. states, collecting over 4 million motion samples, we identify and address four critical deployment challenges previously underexplored in academic settings: (1) false positives from non-human motion sources (pets, robots) that degrade system reliability, (2) hardware heterogeneity in commercial IoT devices causing inconsistent CSI quality, (3) signal interference in multi-user environments limiting individual tracking capabilities, and (4) computational and bandwidth constraints preventing real-time edge processing.
The deployed system integrates a biomechanics-based classifier that reduces non-human false alarms from 63.1\% to 8.4\%, a multi-layer sensing quality metric validating device suitability without environment-specific calibration, proximity-based multi-user detection leveraging distributed IoT devices, and a hybrid edge-cloud architecture with ACF-based compression achieving 99.72\% data reduction. The integrated system achieves 92.61\% human motion detection accuracy across diverse uncontrolled home environments. We have successfully deployed home monitoring technology on millions of WiFi routers nationwide and smart IoT devices (e.g., bulbs, plugs) worldwide, demonstrating its viability for large-scale real-world applications. We share these findings to guide future research toward deployable WiFi sensing systems. 
\end{abstract}

\begin{IEEEkeywords}
Home Monitoring, WiFi Sensing, Internet-of-Things, Edge Devices, Experience
\end{IEEEkeywords}

\section{Introduction}
\label{sec:intro}

\begin{figure}[t]
    \centering
    \includegraphics[width=1\linewidth]{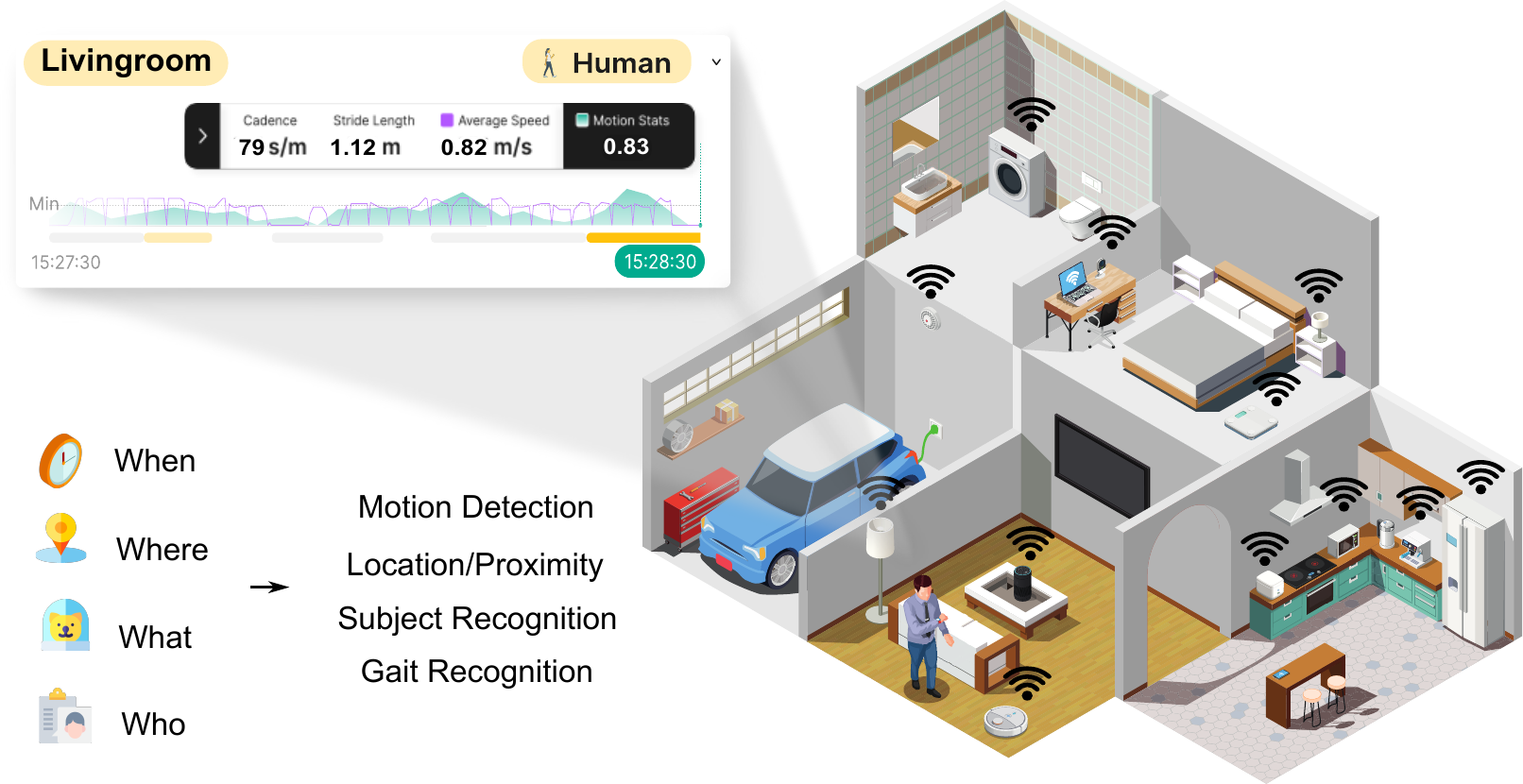}
    \caption{The concept of WiFi-based home monitoring.}
    \label{fig:hm_concept}
\end{figure} 

\IEEEPARstart{H}{ome} monitoring has become increasingly important for individuals seeking to ensure the safety and well-being of their loved ones, protect property against intrusions, and oversee daily routines, particularly for older adults or people with health concerns. The global smart home security market reached USD 29.04 billion in 2024 \cite{grandview_smart_homes_2025,fortune_smart_home_security_2025}, yet existing solutions face fundamental tradeoffs. Traditional solutions, usually based on cameras~\cite{camera_CHI,camera_IPSN,camera_PMC,camera_IEEEP}, wearables~\cite{wearables_JSAC,wearables_TBE,wearables_TITB}, mmwave radars~\cite{mmwave_CST,mmwave_INFOCOM,mmwave_IOTJ}, or PIR sensors~\cite{pir_0,pir_1,pir_2,pir_3}, often demand specialized hardware, complex installations, or intrusive data collection methods, driving the need for systems that operate unobtrusively, preserve user privacy, and integrate seamlessly with existing infrastructure.

While these conventional sensing modalities each offer certain strengths, they share a common limitation: they require dedicated hardware installations that add cost, complexity, and maintenance burden to home environments. PIR sensors, though inexpensive, provide only binary presence detection within line-of-sight of each sensor, requiring multiple units for whole-home coverage and offering no ability to distinguish between humans, pets, or other moving objects. Cameras deliver rich visual information but raise significant privacy concerns, particularly in sensitive areas such as bedrooms and bathrooms, and similarly require line-of-sight coverage. mmWave radars achieve fine-grained sensing but demand specialized hardware installation and careful placement to avoid blind spots.

WiFi-based home monitoring presents a fundamentally different approach: rather than adding new sensing infrastructure, it repurposes the wireless communication signals already permeating modern homes. The key insight is that WiFi signals—continuously transmitted between routers and connected devices—carry embedded information about the physical environment through their multipath propagation characteristics. Human motion perturbs these propagation paths, creating detectable patterns in Channel State Information (CSI) that can be extracted from standard WiFi chipsets without hardware modification.

This approach offers several unique advantages that no single alternative modality can match:
\begin{itemize}
    \item \textbf{Zero additional hardware}: Leverages existing WiFi routers and IoT devices (smart bulbs, plugs, speakers) already present in homes, eliminating installation costs and complexity.
    \item \textbf{Through-wall whole-home coverage}: Unlike PIR sensors and cameras that require line-of-sight, WiFi signals penetrate walls and furniture, enabling a single router-device pair to monitor multiple rooms simultaneously.
    \item \textbf{Privacy by design}: Captures only radio-frequency signal variations and no images, video, or audio, making it inherently suitable for privacy-sensitive spaces.
    \item \textbf{Rich sensing capability}: Beyond binary presence detection, WiFi CSI enables speed estimation, gait analysis, and subject discrimination, supporting applications from intrusion detection to elderly care.
    \item \textbf{Seamless integration}: Operates passively alongside normal WiFi communication without degrading network performance, requiring no user interaction or device wearing.
\end{itemize}

Academic research has demonstrated WiFi sensing's potential across numerous applications~\cite{indotrack_jie,wais_dineash,musefi2023,wifi_can_do_more},
including occupancy detection~\cite{occupancy_1,see_wifi_fadel}, activity recognition~\cite{activity_1,activity_2}, fall detection~\cite{fall_1, fall_2}, breathing detection~\cite{breath_1,breath_2,breath_3}, indoor tracking and mapping~\cite{easitrack,ezmap}, 
human identification~\cite{XModal_ID_Mostofi,gaitway}, and pose estimation~\cite{Yan_2024_CVPR}, etc. 
Although these academic efforts have shown promise, many have been tested in controlled scenarios and limited scales. Accuracy can decline in complex real-world conditions characterized by complex multipath effects, multiple occupants, low-quality IoT chipsets, and environmental dynamics. Therefore, while these preliminary studies lay a theoretical groundwork, bringing them to robust, large-scale deployment has remained an open challenge, requiring practical designs and extensive validation in diverse, everyday settings.

\begin{figure}[t]
    \centering
    \begin{subfigure}[b]{0.2\textwidth}
        \centering
      \includegraphics[height=135pt]{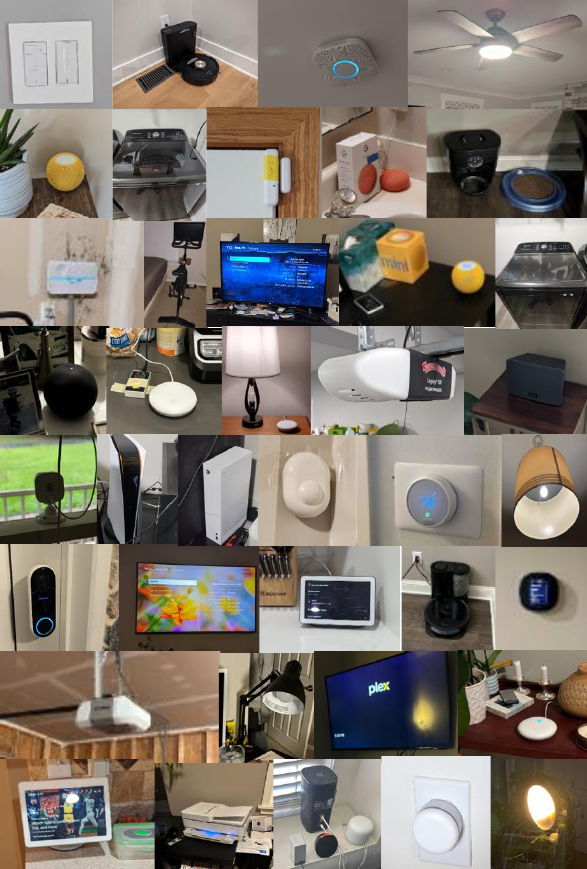}
        \caption{}
        \label{fig:single_office}
    \end{subfigure}%
    \begin{subfigure}[b]{0.08\textwidth}
        \centering
    \includegraphics[height=135pt]{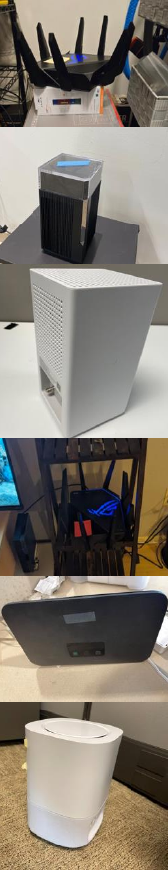}
        \caption{}
        \label{fig:single_home}
    \end{subfigure}%
    \begin{subfigure}[b]{0.2\textwidth}
        \centering
        \includegraphics[height=135pt]{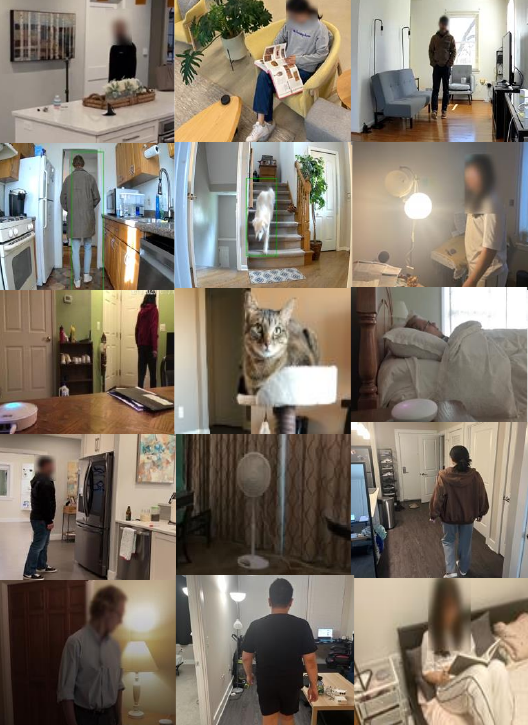}
        \caption{}
        \label{fig:multi_home}
    \end{subfigure}%
    \vspace{-2mm}
    \caption{Deployment in authentic user homes with (a) Existing IoT devices placed by users, (b) Routers located at diverse positions, and (c) Daily activities/motion by varied moving subjects—posing real-world sensing challenges.}
    \label{fig:deplopyment_devices}
\end{figure}

To bridge this gap, we began investigating by deploying existing WiFi sensing systems~\cite{widetect,Wispeed2018,gaitway} on commercial devices. Initial field deployments immediately revealed critical challenges that laboratory studies had not anticipated: commodity hardware with low-quality chipsets varies unpredictably, multipath environments change dynamically, multiple occupants move simultaneously, and system resources are severely constrained. To systematically understand and address these issues, we conducted a comprehensive two-year study across 15 homes in 11 U.S. states, deploying 280 IoT devices to collect over 4 million motion samples—the largest known real-home evaluation of WiFi sensing to date. Fig.~\ref{fig:deplopyment_devices} shows IoT devices and routers deployed in users' homes.

Based on this study, we develop a comprehensive WiFi-based home monitoring system that translates WiFi sensing research into a practical, real-world system. Drawing on extensive commercial experience, in this paper, we share the lessons and insights gained and propose solutions, offering a definitive, affirmative answer to the long-sought question: \emph{Can WiFi sensing be practical and scalable?} 
In fact, the experience and insights shared in this paper enabled our technology to scale in practice. It has been successfully deployed on millions of routers and enables motion sensing on over 100 million smart IoT devices worldwide—demonstrating that WiFi sensing can transition from laboratory research to mass-market commercial products.

The critical challenges emerged from our deployment experience are as follows: 

\noindent\textbf{Challenge 1: Subject Discrimination.}
Non-human motion sources—pets, robotic vacuums, and oscillating appliances—generated false positives in 63.1\% of detection events. Unlike controlled laboratories where only intentional human motion occurs, real homes contain diverse moving entities that trigger identical signal perturbations. These persistent false alarms erode user trust through ``cry wolf'' scenarios, potentially causing genuine security events to be ignored.

\noindent\textbf{Challenge 2: Low-quality CSI.}
Hardware heterogeneity, device placement, and limited coverage affect WiFi sensing reliability. Variability in chipsets, antenna configurations, and manufacturing tolerances leads to inconsistent channel state information (CSI) measurements. Device placements further affect accuracy, as increased distances weaken signals, introduce multipath interference, and lower SNR, thereby reducing motion sensitivity. Additionally, limited coverage and obstacles create blind spots, disrupting continuous, whole-home monitoring. 

\noindent\textbf{Challenge 3: Multi-User Ambiguity.}
WiFi's limited spatial resolution cannot disambiguate overlapping signals when multiple occupants move simultaneously. While single-user scenarios dominate academic evaluations, typical households have multiple residents whose concurrent movements create signal interference patterns that standard algorithms cannot separate, limiting applications requiring individual tracking (\textit{e.g.}, healthcare and security) where ambiguous or incorrect identification can undermine system effectiveness.

\noindent\textbf{Challenge 4: Resource Constraints.}
Edge routers cannot execute complex machine learning models while maintaining primary networking functions—our hardware partners limit sensing to 7\% CPU utilization. Yet streaming raw CSI to the cloud incurs a burden on residential uplinks and raises privacy concerns. This creates a fundamental tension between computational requirements and available resources.

Beyond these core challenges, we investigate practical deployment considerations including system architecture and the interaction between sensing and communication functions. Specifically, we examine whether WiFi sensing degrades network performance for latency-sensitive applications like video calling—a critical concern for consumer acceptance.

Through iterative development and validation in our testbed homes, we designed practical solutions for each challenge. Our integrated system achieves 92.61\% human motion detection accuracy across diverse uncontrolled environments, enabling solutions successful commercialization. 
By sharing the key challenges and the lessons learned, we provide practical guidance for making WiFi sensing viable in real-world settings.
In this paper, we primarily focus on home monitoring, one of the killer applications of WiFi sensing. 
Yet the methods and experiences shared here are broadly applicable to various WiFi-based applications, from health monitoring to smart buildings. Ultimately, our work paves the way for more widespread adoption of WiFi sensing technologies, shaping the world into a smarter, safer living space.

In the rest of the paper, we will present related work in \S\ref{sec:related_work} and the system design in \S\ref{sec:homemonitoring}. We address the four challenges in \S\ref{sec:sub}, \S\ref{sec:quality}, \S\ref{sec:multi_user}, and \S\ref{sec:computation}, respectively. We introduce the deployment architecture of our system in \S\ref{sec:deploy_archit} and discuss the mutual impact of WiFi sensing and communication in \S\ref{sec:ISAC}. We conclude the paper in \S\ref{sec:con}.

\begin{table*}[t]
\centering
\caption{
Comparison of \textit{alternative} sensing modalities for home monitoring. This table surveys existing technologies that could be used for home monitoring; our proposed system uses \textit{WiFi sensing only} and does not incorporate these other modalities. The ``Pricing Info'' column shows example commodity hardware costs. Columns \textit{Contactless} show 
\cmark\ for ``Yes'' and \xmark\ for ``No.'' Other columns use text-based ratings (High, Moderate, Low) or descriptive terms.%
}
 \small
\begin{tabularx}{\textwidth}{
    C{0.11\textwidth}|  
    C{0.07\textwidth}    
    C{0.07\textwidth}   
    C{0.07\textwidth}   
    C{0.15\textwidth}   
    C{0.10\textwidth}   
    C{0.12\textwidth}   
    C{0.12\textwidth}   
    C{0.085\textwidth}  
}
\toprule
\textbf{Modality} & \textbf{Image} & \textbf{Privacy} & \textbf{Contactless} & \textbf{Pricing Info} & \textbf{Coverage} & \textbf{Adoption Rate} & \textbf{User Acceptance} & \textbf{Resolution} \\
\midrule

\textbf{WiFi (Ours)} &
\includegraphics[width=0.03\textwidth]{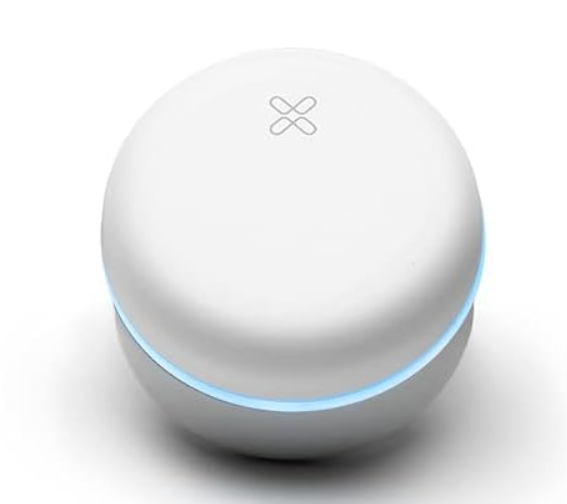} &
High &
\cmark & 
\$0, reuses existing devices &
Whole Home &
High & 
High &
Moderate \\
\midrule

Cameras  &
\includegraphics[width=0.03\textwidth]{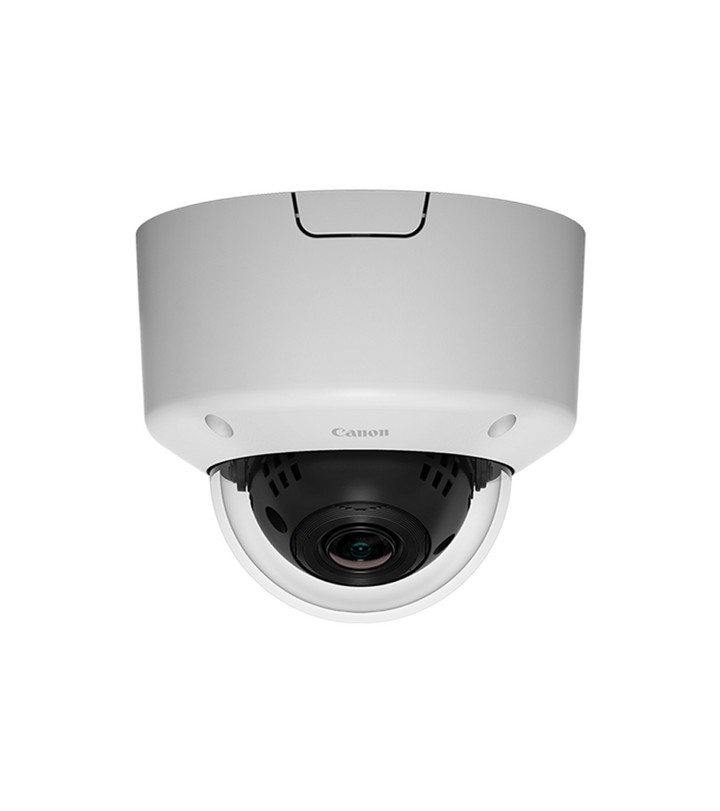} &
Low &
\cmark &
\$35, Waze Cam&
Line of Sight &
Medium & 
Moderate &
High \\
\midrule

PIR Sensors &
\includegraphics[width=0.03\textwidth]{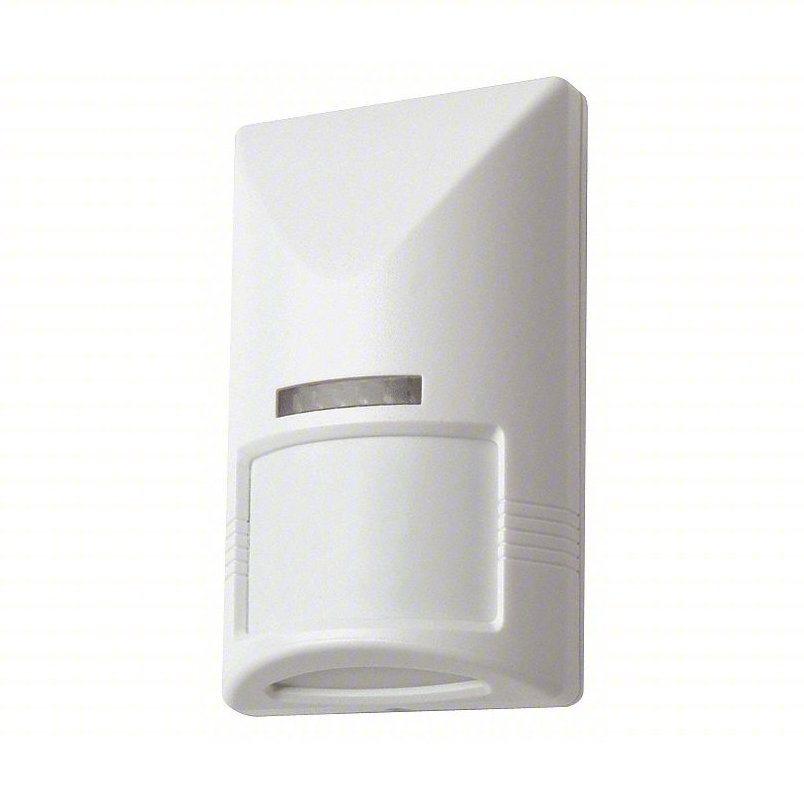} &
High &
\cmark &
\$10, Panasonic EKMC &
Room &
Medium &
High &
Low \\
\midrule

Wearables &
\includegraphics[width=0.03\textwidth]{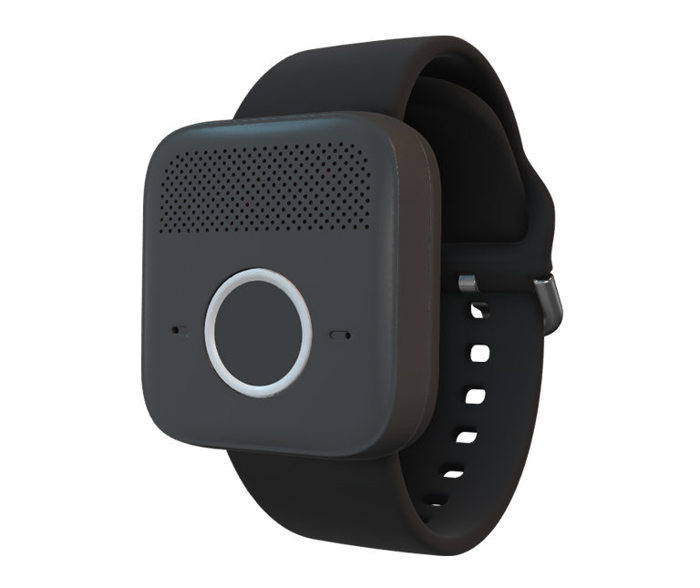} &
Moderate &
\xmark &
\$99, Fitbit&
Individual &
Medium &
Low &
High \\
\midrule

mmWave Radar &
\includegraphics[width=0.03\textwidth]{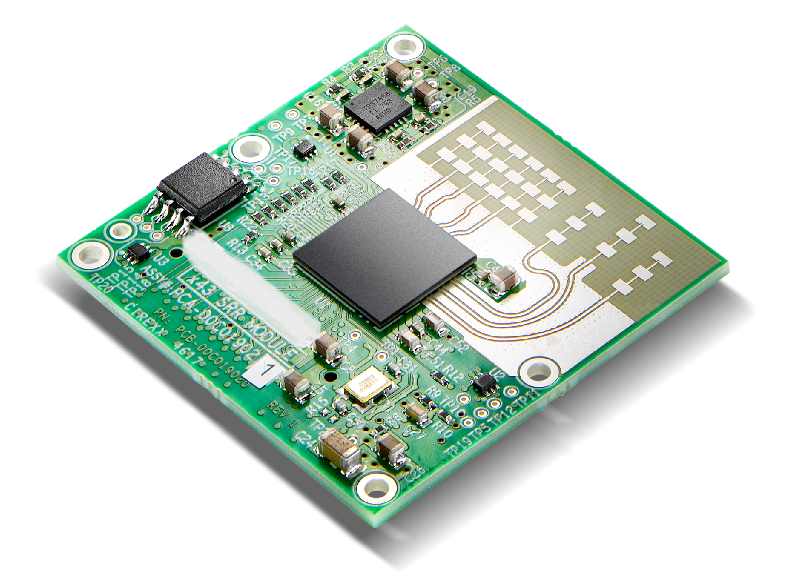} &
High &
\cmark &
\$20, Infineon BGT60&
Room &
Low &
Moderate &
High \\
\bottomrule
\end{tabularx}
\label{tab:ModalityComparison}
\end{table*}

\section{Related Work}
\label{sec:related_work}

This section surveys related work across conventional sensing modalities for home monitoring and WiFi-based sensing research.

\subsection{Conventional Sensing Modalities}

Table~\ref{tab:ModalityComparison} compares popular sensing modalities for home monitoring. Each modality represents an \textit{alternative approach} to the home monitoring problem, with distinct hardware requirements, coverage characteristics, and privacy implications. Our WiFi-based system does not incorporate or fuse data from these alternative modalities; rather, we position WiFi sensing as a complementary or replacement technology that addresses their limitations.

\textbf{Camera-based systems} provide rich visual information enabling detailed activity recognition and person identification~\cite{camera_CHI,camera_IPSN,camera_PMC,camera_IEEEP}. However, they raise significant privacy concerns~\cite{sound_privacy,camera_privacy_concern1,camera_privacy_concern2}, require line-of-sight coverage, and users often disable them in private spaces such as bedrooms and bathrooms~\cite{camera_privacy_concern3,camera_privacy_concern4}. Recent work has explored privacy-preserving approaches using silhouettes or skeleton extraction, but these still require camera installation throughout the home.

\textbf{PIR sensors} detect thermal radiation changes from human presence and are widely deployed due to low cost (\$10--15 per unit)~\cite{pir_0,pir_1,pir_2,pir_3}. However, they provide only binary presence detection (occupied/unoccupied) within their line-of-sight cone, cannot distinguish between humans and pets, and are susceptible to false alarms from temperature fluctuations and airflow~\cite{pir_false_alarm}. Achieving whole-home coverage requires installing multiple sensors in each room, adding deployment complexity.

\textbf{mmWave radars} achieve fine-grained motion tracking and vital sign monitoring~\cite{mmwave_CST,mmwave_INFOCOM,mmwave_IOTJ}, but require dedicated hardware installation (\$20--100+ per unit), careful placement for coverage optimization, and generally operate within single-room line-of-sight ranges. While commercial products like Google Nest Hub use mmWave for sleep tracking, scaling to whole-home monitoring remains cost-prohibitive.

\textbf{Wearable devices} enable continuous health monitoring with high accuracy~\cite{wearables_JSAC,wearables_TBE,wearables_TITB}, but require user compliance in wearing and charging devices~\cite{wearables_less_use1,wearables_less_use2}. Studies show that elderly users—a primary target demographic for home monitoring—frequently forget or refuse to wear such devices, limiting practical effectiveness.

\subsection{WiFi-Based Sensing}

WiFi sensing has emerged as a promising contactless sensing paradigm that repurposes existing wireless infrastructure. Research spans multiple application domains:

\textbf{Motion and occupancy detection.} Early work demonstrated that WiFi signal variations can detect human presence~\cite{see_wifi_fadel,widetect}. WiDetect~\cite{widetect} introduced a calibration-free approach using statistical CSI features, achieving robust detection across diverse environments. However, these systems do not distinguish human motion from other moving sources.

\textbf{Activity recognition.} Systems like WiFall~\cite{fall_1}, E-eyes~\cite{e-eye}, CARM~\cite{activity_1}, and WiChase~\cite{activity_2} recognize specific activities from CSI patterns. While achieving high accuracy in controlled settings, they typically require environment-specific training data and struggle with cross-domain generalization.

\textbf{Localization and tracking.} IndoTrack~\cite{indotrack_jie}, EasiTrack~\cite{easitrack}, and similar systems enable device-free indoor tracking. Recent work like WAIS~\cite{wais_dineash} integrates WiFi sensing with SLAM for resource-efficient mapping. These approaches demonstrate WiFi's potential for spatial awareness but often require multiple access points or specialized antenna configurations.

\textbf{Vital sign monitoring.} WiFi-based breathing~\cite{breath_1,breath_2,breath_3} and heart rate detection have shown clinical-grade accuracy in controlled conditions. SMARS~\cite{smars} demonstrated sleep monitoring through walls, highlighting WiFi's unique through-wall sensing capability.

\textbf{Person identification.} GaitWay~\cite{gaitway} and XModal-ID~\cite{XModal_ID_Mostofi} identify individuals through gait patterns extracted from WiFi signals. These systems achieve 90\%+ accuracy with enrolled users but require initial training phases.

\textbf{Multi-user sensing.} Recent work addresses the challenging multi-user scenario. MUSE-Fi~\cite{musefi2023} exploits near-field channel variations to separate multiple users, while WiMANS~\cite{huang2024wimansbenchmarkdatasetwifibased} provides benchmark datasets for multi-user activity sensing. Person-in-WiFi 3D~\cite{Yan_2024_CVPR} demonstrates end-to-end multi-person 3D pose estimation. However, these approaches have not been validated at scale in real home environments.

\textbf{Relationship to our prior work.} WI-MoID and its conference precursor develop an interpretable method for human and non-human motion recognition~\cite{wi_moid,Guozhen_conference_version}, while SrcSense studies a complementary deep-learning formulation~\cite{SrcSense}. These studies establish the underlying motion-source recognition techniques under smaller-scale evaluations. The present paper does not claim these techniques as new; it focuses on building on this technical foundation in a complete home-monitoring system and evaluating the resulting system at deployment scale.

Despite this rich body of research, a significant gap remains between laboratory demonstrations and practical deployment. Most studies use research-grade hardware (Intel 5300 NICs, Atheros CSI tools), controlled environments with known layouts, and limited participant pools. Our work bridges this gap by validating WiFi sensing across 15 real homes with commodity IoT devices, identifying and solving challenges that emerge only at deployment scale.

We position WiFi sensing as a practical, scalable foundation for intelligent home monitoring. While we focus on the sensing layer, the rich motion, speed, and presence information extracted enables downstream applications in knowledge reasoning and intelligent control.

\section{WiFi-based Home Monitoring System}
\label{sec:homemonitoring}

This section presents our WiFi-based home monitoring system design. We first describe the sensing principle and baseline foundation module, then overview the complete system architecture that addresses the deployment challenges introduced in \S\ref{sec:intro}.

\subsection{Sensing Principle and Foundation Module}
\label{sec:system_design}

Our system extracts motion information from Channel State Information (CSI)—the fine-grained channel frequency response measured by WiFi chipsets during normal packet reception. Unlike received signal strength indicator (RSSI), which provides only a single aggregate power measurement, CSI captures amplitude and phase across multiple subcarriers (typically 56--256 depending on bandwidth), enabling detection of subtle motion-induced signal variations.

Human movement perturbs the multipath propagation environment and creates temporal correlations in CSI. The foundation module used in our system summarizes established ACF-based motion-detection and speed-estimation methods from prior work~\cite{widetect,Wispeed2018,gaitway,wi_moid,SrcSense}. We retain the notation needed to explain its role in the deployed system; complete derivations and algorithmic validation are provided in the cited papers. This module is not a new algorithmic contribution of the present work.

For completeness, we briefly summarize the foundation used by the deployed system. In contrast to geometry-based reflection models, the module adopts a statistical electromagnetic scattering model and characterizes motion-induced temporal correlation through the autocorrelation function (ACF):

\begin{equation}
    \rho_G(\tau, f) = \frac{\mathbb{E}[G(t, f) G(t - \tau, f)] - \mu^2(f)}{\sigma^2(f)},
\end{equation}
where $G(t,f)=|H(t,f)|^2$ is the CSI power and $\tau$ is the time lag. In static environments, $\rho_G(\tau,f)$ is close to zero for $\tau \ne 0$, while motion increases the temporal correlation. A motion statistic $\phi(f)=\rho_G(\Delta t,f)$ is compared with a globally determined threshold, enabling calibration-free motion detection without site-specific tuning~\cite{widetect}.

For through-wall sensing, the system aggregates motion-induced changes across multipath components. Under the established scattering model, the aggregated ACF is represented as
\begin{equation}
    \rho_G(\tau) = \frac{1}{F} \sum_{f=1}^{F}
    \frac{E_d^2(f)}{E_d^2(f) + \sigma^2(f)} J_0(kv\tau),
\end{equation}
where $E_d(f)$ is the motion-induced energy, $J_0$ is the zeroth-order Bessel function, $k=2\pi/\lambda$ is the wavenumber, and $v$ is the target speed. The speed is estimated from the first peak $\tau_{\text{peak}}$ of the differential ACF:
\begin{equation}
    v = \frac{x_0 \lambda}{2\pi \tau_{\text{peak}}},
\end{equation}
where $x_0$ is the first peak of $J_0$. Maximal ratio combining is then applied across subcarriers to improve robustness in low-SNR conditions, achieving an 8--12 dB SNR gain and enabling through-wall motion and speed detection at distances exceeding 10~m~\cite{smars}.

As established in the cited foundation works, multipath can be exploited as a statistical advantage: more reflected paths improve stability, with $\mathrm{Var}(\rho_G(\tau)) \propto 1/N$~\cite{Hill2009}. Using only CSI amplitude $|H(t,f)|^2$ also makes the module \textit{phase-agnostic}, reducing sensitivity to carrier and sampling frequency offsets (CFO/SFO) in commodity chipsets~\cite{xie2015precise,Chen2017}. These properties ensure compatibility and robustness across diverse hardware types and deployment settings.

\begin{figure}[t]
    \centering
    \includegraphics[width=0.95\linewidth]{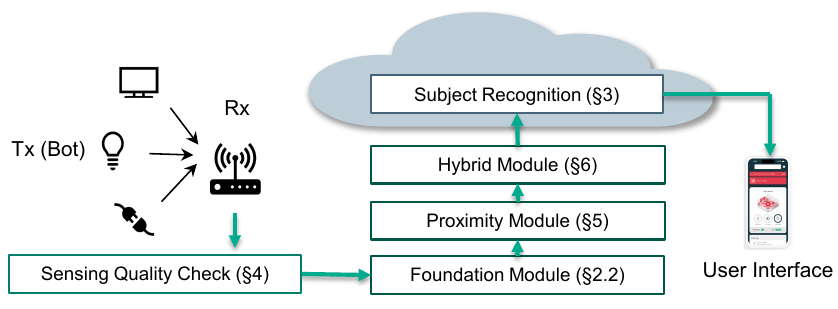}
        \vspace{-2mm}
    \caption{Overview of WiFi-based home monitoring.}
    \label{fig:sys-overview}
\end{figure}

\subsection{From Foundation to Practical System}

While the foundation module offers reliable motion detection in controlled settings, extensive real-world deployment revealed several limitations that hinder its direct scalability. Commodity IoT devices exhibit inconsistent CSI quality, non-human movements such as pets and robots cause false positives, multiple occupants produce overlapping signals, and edge devices face strict computation and bandwidth constraints. These challenges collectively motivated a series of practical system-level solutions.

Building on the baseline module, we design a modular and hierarchical WiFi-based home monitoring system, illustrated in Fig.~\ref{fig:sys-overview}. This architecture integrates several key components:
\begin{itemize}
    \item \textbf{Subject discrimination} (\S\ref{sec:sub}) identifies human motion based on biomechanics-inspired features.
    \item \textbf{Sensing-quality verification} (\S\ref{sec:quality}) assesses CSI stability and device suitability before activation.
    \item \textbf{Proximity-based multi-user sensing} (\S\ref{sec:multi_user}) leverages distributed IoT devices to localize concurrent occupants.
    \item \textbf{Hybrid edge--cloud computing} (\S\ref{sec:computation}) balances real-time responsiveness and scalable processing under constrained resources.
\end{itemize}

On the \textbf{edge side}, high-quality CSI is captured, filtered, and processed by the foundation module, which computes ACF-based motion and speed features in real time. A hybrid controller then determines whether to offload processed ACF data to the cloud for further analysis.  
On the \textbf{cloud side}, the sensing server performs subject classification, fuses data from multiple devices, and updates user interfaces with motion events, device health, and system statistics.

Together, these modules transform the baseline foundation into a practical home monitoring system that achieves robust, privacy-preserving, and scalable performance across millions of deployed WiFi and IoT devices. 

Our field deployment spans 15 residential homes across 11 U.S. states, covering a diverse mix of physical layouts and device environments. Participating homes range from compact apartments (600–800 ft²) to multi-floor single-family houses (2,500+ ft²). Each home is equipped with 5 to 30 WiFi-enabled IoT devices, such as smart lights, plugs, routers, and hubs, operating over both 2.4 GHz and 5 GHz bands. Devices are deployed in a variety of locations to capture both line-of-sight (LOS) and non-line-of-sight (NLOS) propagation conditions, reflecting realistic indoor sensing challenges. The deployment includes multiple commodity chipset vendors, including Qualcomm, MediaTek, Realtek, and Espressif, spanning different CSI extraction capabilities and radio behaviors. These varied configurations enable us to evaluate sensing robustness under diverse real-world conditions.

The following sections (\S\ref{sec:sub}--\S\ref{sec:computation}) describe the key deployment challenges uncovered during large-scale testing and our solutions to each of them.

\begin{figure}[t]
    \centering
    \includegraphics[width=1\linewidth]{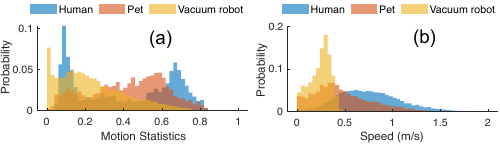}
        \vspace{-4mm}
    \caption{  Histograms of (a) motion statistics and (b) speed estimations of humans, pets, and robots.}
    \label{fig:sys_hist_speed_ms}
\end{figure}

\begin{figure}[t]
    \centering
    \includegraphics[width=1\linewidth]{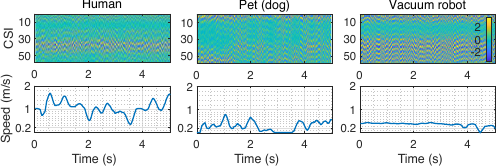}
        \vspace{-4mm}
    \caption{  Distinct speed patterns of humans, pets, and vacuum robots.}
    \label{fig:speed_pattern}
\end{figure}

\section{Subject identification}
\label{sec:sub}
In practical home monitoring scenarios, various subjects including humans, pets, robotic cleaners, and oscillating or rotating electrical appliances can interfere with WiFi signals. As a result, identifying the subject of interest is crucial for WiFi-based home monitoring. Applications demanding high specificity—such as intruder detection, energy optimization, and health monitoring—cannot function effectively if every non-human movement triggers an undesired alert.

\head{Challenges.}
Although our foundational module excels in detecting general motion, it struggles to discriminate between humans and non-human moving subjects in real-world environments. In realistic home environments, these unrelated subjects frequently induce false detections, eroding system reliability. These subjects reflect or scatter WiFi signals, generating similar MS to humans, as shown in Fig. \ref{fig:sys_hist_speed_ms}. Existing solutions, such as PetFree~\cite{petfree}, rely on rigid device placement and assumptions ill-suited for diverse home layouts. 

We initially attempted to suppress false alarms using fixed motion-intensity thresholds derived from amplitude dynamics. However, this approach proved unreliable: small pets in close proximity to Bots often produced signal variations indistinguishable from those caused by humans, while slow-moving individuals were occasionally missed. We further explored filtering based on motion speed alone, but this too fell short—speed profiles of humans and pets overlapped significantly. For example, a cat's rapid dash or a robot vacuum's abrupt movement can mimic the MS of human activity, while a slow-moving adult may generate a similar signature to a pet strolling nearby (Fig. \ref{fig:sys_hist_speed_ms}b). As shown in Fig.~\ref{fig:sys_hist_speed_ms}(b), the speed distributions of humans and pets exhibit substantial overlap, particularly in the 0.5--1.5 m/s range, confirming that speed magnitude alone cannot reliably distinguish between subject types in real-world environments. These limitations motivated our shift toward a biomechanics-informed classifier that considers both physical and statistical features beyond simple intensity or speed measures.

\begin{figure}[t]
    \centering
    \includegraphics[width=0.9\linewidth]{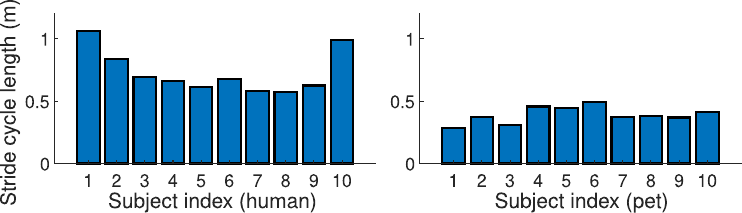}
    \caption{Stride length difference of humans and pets.}
    \label{fig:gait_cycle}
\end{figure}

\begin{figure}[t]
    \centering
    \includegraphics[width=0.9\linewidth]{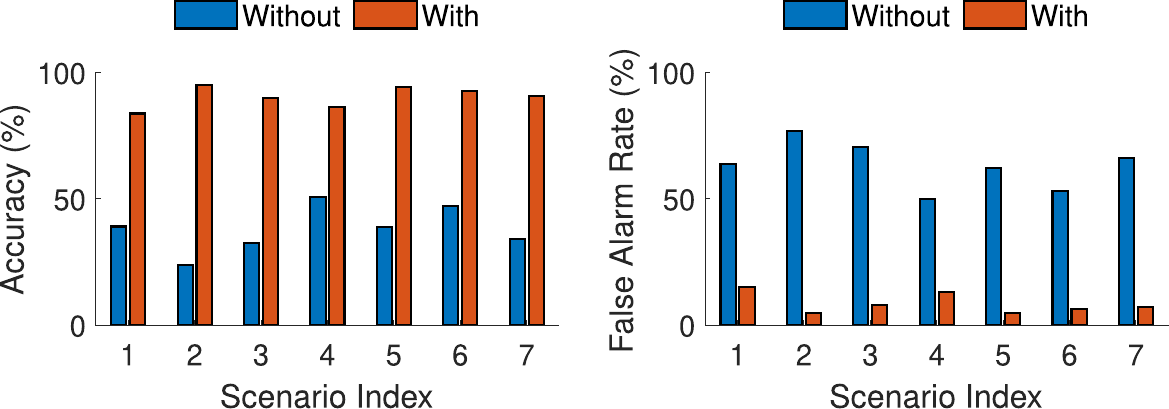}
    \caption{Comparison of accuracy and false alarm rate with and without subject-of-interest recognition.}
    \label{fig:subject_recg_results}
\end{figure}

\head{Approaches.}
The deployed subject-identification module builds on the biomechanics-based motion-source recognition method introduced in WI-MoID and its conference precursor~\cite{wi_moid,Guozhen_conference_version}; SrcSense presents a complementary deep-learning formulation~\cite{SrcSense}. Here, we focus on deployment-scale evaluation and system integration rather than claiming a new motion-source recognition algorithm.

The classifier exploits differences in movement mechanics: humans exhibit \textit{bipedal} stance--swing cycles, pets move \textit{quadrupedally}, and wheeled robots maintain relatively consistent speeds. Fig.~\ref{fig:speed_pattern} shows representative CSI and speed patterns for a human, a dog, and an iRobot. Because instantaneous speeds can overlap across subject types (Fig.~\ref{fig:sys_hist_speed_ms}b), the classifier uses temporal dynamics and periodicity, including stride length and cycle time (Fig.~\ref{fig:gait_cycle}), together with the smoother motion profiles of wheeled robots.

Building on this prior framework, the deployed classifier analyzes speed patterns and signal statistics using seven physical features---including gait existence, stride length, cycle time, average speed, speed variance, and speed percentiles---and six statistical features, including ACF peak/valley means, peak/valley intervals, and motion-statistic mean/variance. The resulting 13-dimensional vector is classified by an SVM. This section evaluates the module across diverse homes using leave-one-home-out validation and studies its role in the complete monitoring system.

\head{Experiments.}
To evaluate the deployed module at scale, we collect over 2.4 million data samples spanning 84 days in 7 homes, from small apartments to single-family houses, with each sample representing a 6-second window. We label motion data for non-human subjects in situations where no humans are present, aiming to distinguish such events from human motion. During testing, we make no model adjustments or classifier retraining for any unseen environment. Our dataset captures everyday human activities, pets of varying sizes (5 dogs and 4 cats ranging from 7.5 lb to 50 lb), robotic vacuums under normal household use, and oscillating or ceiling fans. All subjects move freely, often behind walls or furniture, ensuring realistic conditions. Importantly, the dataset includes challenging cases where speed profiles overlap significantly, such as slowly walking elderly residents, energetic small dogs, and fast-moving robot vacuums, reflecting the real-world difficulty highlighted by existing literature.

\begin{figure*}[t]
    \centering
    \includegraphics[width=0.98\linewidth]{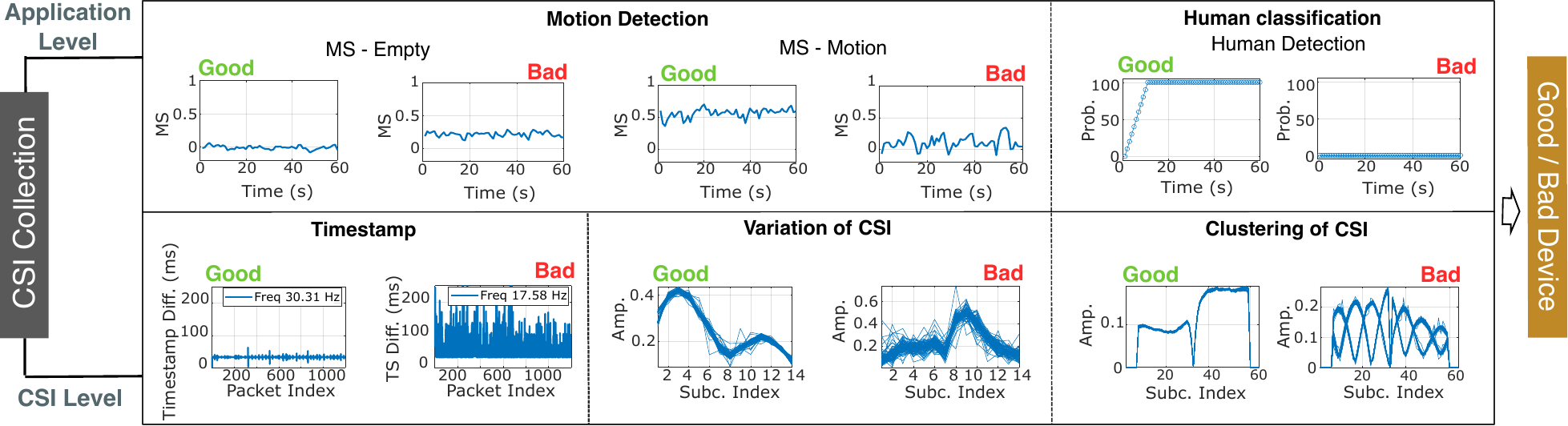}
    \caption{  The framework of the sensing verification~\cite{csiverification} for evaluating the sensing quality of different edge devices. 
    }
    \label{fig:diagram_CSIveri}
\end{figure*}

To assess robustness, we apply a leave-one-environment-out method, training the SVM classifier on data from 6 environments and evaluating it in the remaining unseen environment. The testing results in Fig. \ref{fig:subject_recg_results} show that our approach lowers the average false-alarm rate for non-human subjects from 63.1\% to 8.4\% and increases the average recognition accuracy for human vs. non-human motion from 37.9\% to 90.4\%, even under NLOS conditions. These gains highlight the value of biomechanical cues over purely heuristic-based methods and confirm that the distinct CSI signals generated by bipedal strides, quadrupedal gaits, and continuous wheel rotations can be effectively separated. Additionally, because our method does not rely on environment-specific training data or rigid device placement, it adapts well to everyday living spaces.

To further reduce false alarms, we implement a simple temporal confidence scoring on top of the SVM output. Instead of acting on a single classification, we aggregate decisions over a short time window. Specifically, we fuse outputs from multiple receivers (if available) and use a sliding window to ensure consistent classification before confirming a human detection. Transient misclassifications are outweighed by consistent votes from others over time. We normalize the aggregated score to a 0–99 range and only trigger a human-detected alert if the score exceeds a threshold. This method provides resilience to sporadic errors and gives users more confidence by presenting a real-time ``confidence level'' for each detection in the UI.

\head{Findings.}
\emph{Focusing on underlying physical locomotion signatures—rather than isolating a single metric such as speed—transforms WiFi-based sensing into a robust, physics-informed tool for real-world applications.} By modeling the nature of human, pet, and robotic motion, our framework mitigates interference from non-human activities, enhancing reliability in actual households and laying the foundation for user-centric, trustworthy smart environments. We also observe that when humans and pets move together in the same area, the human motion profile dominates, enabling the system to consistently detect human motion.

\section{Sensing quality with IoT devices}
\label{sec:quality}
A survey by Reviews.org~\cite{reviews2024smarthome} found that 85\% of Americans own at least two smart home IoT devices. A home monitoring system can harness existing WiFi-enabled IoT devices and a central router for environmental sensing. 
The variety of these devices, however, poses new challenges. 

\subsection{Challenges}

\noindent\textbf{Hardware Heterogeneity.} Variations in chipsets, signal processing, and antenna configurations can affect CSI reliability—even among identical models~\cite{practicalissue2022, hardwarerelated}. Robust evaluation is essential to confirm if a device can serve effectively as a sensing tool.

\noindent\textbf{Device Placement.} Environmental factors and placement significantly impact WiFi sensing accuracy~\cite{practicalissue2022, daqing2023placement, wallmatters}. Increased distances or obstacles weaken signals and reduce motion sensitivity, highlighting the need for performance assessments across diverse settings.

\noindent\textbf{Limited Coverage.} A single wireless link is constrained by range, signal loss, and obstacles that create blind spots~\cite{practicalissue2022, wallmatters}. Evaluating device performance under these conditions is critical for ensuring reliable tracking and connectivity.

\subsection{Sensing Quality of Diverse Devices}

\begin{figure*}[t]
    \centering
    \includegraphics[width=0.96\linewidth]{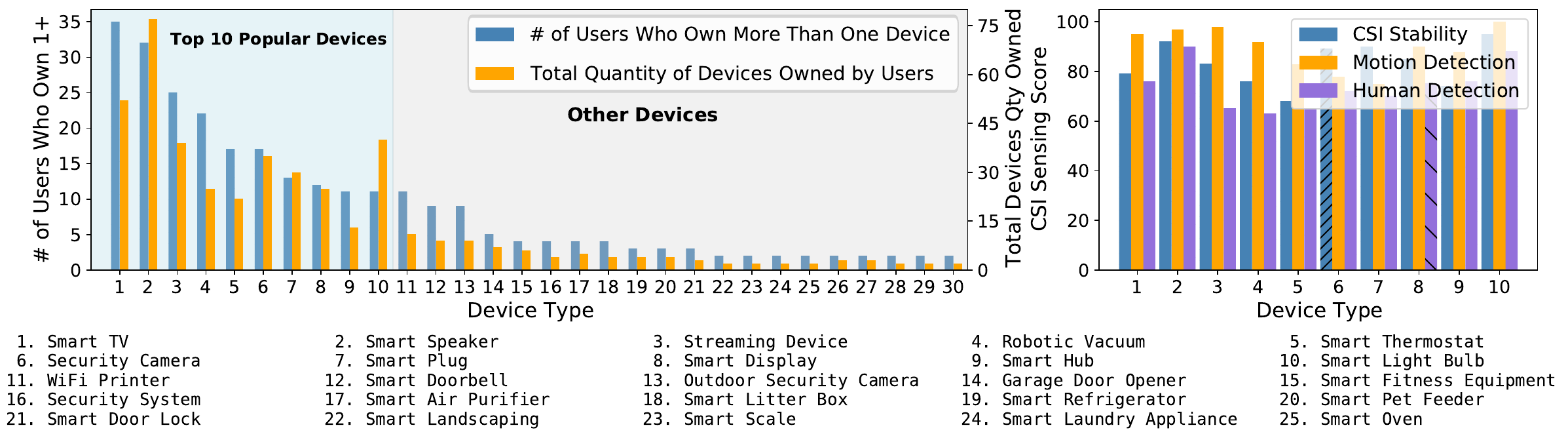}
    \vspace{-3mm}
    \caption{  The survey of the IoT device popularity (left) and the CSI scores for different types of IoT devices (right).}
    \label{fig:iot_score}
\end{figure*}

\head{Approaches.} Robust evaluation of sensing quality requires a well-defined quantitative criterion. Prior work provides SSNR-based sensing-capability measures~\cite{daqing2023placement} and CSI-level metrics for packet timestamp consistency and amplitude stability~\cite{csiverification}. Building on the CSI-level metrics in~\cite{csiverification}, we extend the sensing-quality framework with application-level motion-detection and human-classification measures tailored to home monitoring (see Fig. \ref{fig:diagram_CSIveri}). Separate statistical measures for packet timestamps and CSI amplitude are normalized within 
\(0\)--\(100\), and the final sensing quality score is computed as a weighted average of both CSI-level and application-level scores.

To assess a device's sensing capability, users complete a one-minute calibration process: walk near the edge device for 30 seconds, then remain stationary for another 30 seconds while CSI data is collected and analyzed. Analysis of extensive data from diverse deployments and device models—including various house types—demonstrates that when the final score exceeds 60, the device is generally qualified for sensing by consistently maintaining a packet loss rate below 10\% and achieving 80\% accuracy in both motion detection and human classification. 

To improve usability, we deploy the sensing verification algorithm as a pre-screening module on edge devices. Before activating the sensing application, users can run the qualification test to assess device suitability based on sensing quality scores. Devices that do not meet the threshold can have their sensing functions disabled.

\head{Experiments.} To analyze sensing quality across popular smart devices, we collect CSI from IoT devices used by 40 users and conduct a survey on device ownership (left figure, Fig. \ref{fig:iot_score}). Based on the results, we select the 10 most popular device types, including smart TVs, speakers, streaming devices, robotic vacuums, thermostats, security cameras, smart plugs, displays, hubs, and light bulbs. 
Devices span major chipset vendors such as Qualcomm, Broadcom, Espressif, and MediaTek. Table \ref{tab:device_summary} lists examples of their specifications. We perform sensing quality checks to quantitatively evaluate the signal stability and sensing performance of these devices. 

\begin{table}[h]
\centering

\caption{Summary of edge devices, example WiFi chipsets and models.}
\label{tab:device_summary}
\small
\begin{tabular}{>{\arraybackslash}m{2.2cm} >{\raggedright\arraybackslash}m{2.7cm} >{\raggedright\arraybackslash}m{2.9cm}}
\toprule
\textbf{Device} & \textbf{WiFi Chipset Vendor} & \textbf{Model} \\
\midrule
\multirow{2}{=}{Smart TV} & Realtek & RTL8821CS \\
                          & Broadcom & BCM4356/BCM43569  \\
\multirow{2}{=}{SmartPlug} & MediaTek & MT7697N \\
                           & Espressif & ESP8266/ESP8285 \\
\multirow{2}{=}{Smart Display} & MediaTek & MT8516 \\
                               & MediaTek & MT8183 \\
\multirow{2}{=}{Smart Speaker} & Qualcomm & IPQ4019 \\
                               & MediaTek & MT6625L/MT7658CSN \\
Smart Hub & Broadcom & BCM4345 \\
\multirow{2}{=}{Robot Vacuum} & Qualcomm & WCN3620 \\
                              & Espressif & ESP8266 \\
\multirow{2}{=}{Smart Thermostat} & Texas Instruments & CC3220SF \\
                                  & Broadcom & BCM43362 \\
\multirow{2}{=}{Streaming Device} & MediaTek & MT8695 \\
                                  & Broadcom & BCM43570 \\
\multirow{2}{=}{Smart Light Bulb} & Espressif & ESP8266 \\
                                  & Realtek & RTL8710BN \\
\multirow{2}{=}{Security Camera} & HiSilicon & Hi3518EV300\\
                                       & MediaTek & MT7688 \\
\bottomrule
\end{tabular}
\end{table}

The right figure of Fig. \ref{fig:iot_score} presents sensing scores across three metrics: CSI Stability (blue), Motion Detection (orange), and Human Detection (purple). Scores are averaged across multiple devices within each category. Smart hubs and light bulbs achieve high motion detection scores, while streaming devices and robotic vacuums exhibit lower CSI stability and human detection scores. Human detection generally scores lower due to stricter conditions. 

\head{Findings.}
\emph{While both chipset model and device type influence sensing quality, our results suggest that the chipset model plays a more significant role.} Even within the same device category, CSI stability varies across different chipsets, highlighting the impact of hardware design and signal processing techniques. 
Among the top 10 most popular commercially available IoT devices, every device achieves a score above 60—qualifying them as effective sensing devices for home monitoring. By collecting just one minute of CSI data during calibration, users can rapidly assess each device's suitability through real-time feedback from our sensing verification tool, thereby streamlining the selection process with minimal effort.

\subsection{Device Placement}

\begin{figure}[t]
    \centering
    \includegraphics[width=1\linewidth]{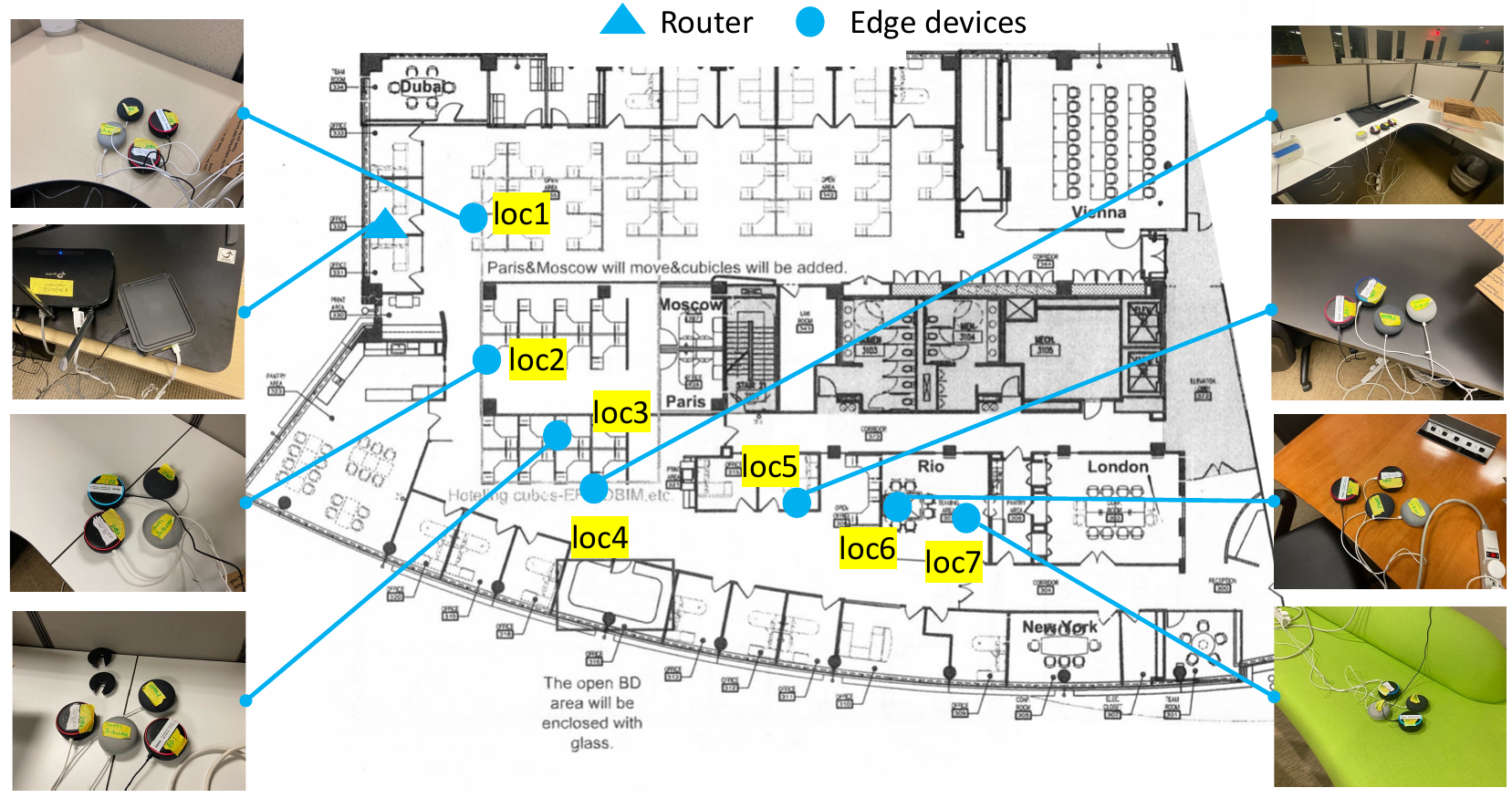}
    \caption{Office testbed layout used for controlled distance-sweep experiments.}
     
    \label{fig:office_layout}
\end{figure}

\head{Experiments.}Apart from the inherent characteristics of the chipset, device placement significantly impacts sensing quality~\cite{daqing2023placement}. While previous work examines how transmitter--receiver deployment relative to walls affects sensing performance~\cite{wallmatters}, we focus on the router--edge IoT device distance as a key factor. 
To isolate this variable under controlled conditions, we conduct a distance-sweep experiment in a large office testbed (Fig.~\ref{fig:office_layout}). We emphasize that this office environment is used \emph{only} for controlled placement testing (not as a representative home deployment): it provides a larger continuous area to realize a wide range of link distances, and it enables us to minimize confounding factors such as uncontrolled human activity and external interference. 
Specifically, we place the edge device at Locations \#1--\#7, where Location \#1 is closest and Location \#7 is farthest from the router. The corresponding router--device distances are 3\,m, 5\,m, 6.5\,m, 10.8\,m, 15\,m, 17\,m, and 18.7\,m, respectively. During this experiment, the environment is kept static (no other moving objects) and we ensure no other significant wireless interference, so that the observed sensing degradation can be primarily attributed to distance.

\begin{figure*}[t]
    \centering
    \includegraphics[width=1\linewidth]{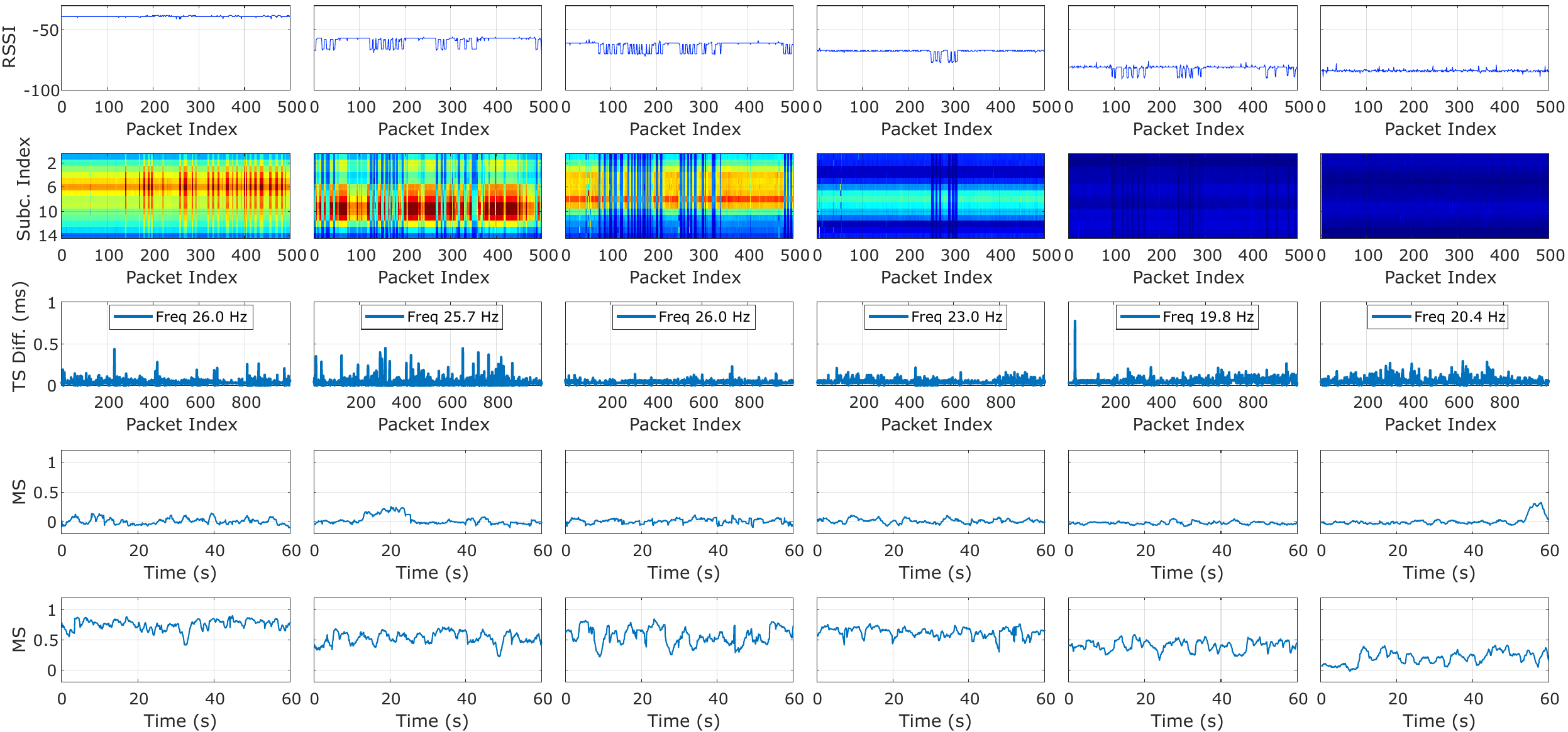}
    \vspace{-6mm}
    \caption{Sensing characteristics at various distances. Rows from top to bottom represent: RSSI, CSI amplitude, timestamp (TS) differential, motion statistics (MS) in static scenario, and MS in dynamic scenario. Columns from left to right correspond to Locations \#1 to \#6 (from near to far), respectively.}
    \label{fig:distance_performance}
\end{figure*}

\head{Findings.}
To analyze sensing quality, we examine multiple signal properties as shown in Fig. \ref{fig:distance_performance}. 
From top to bottom, the figure presents RSSI, CSI amplitude, timestamp (TS) differential, motion statistics (MS) in an empty scenario, and motion statistics (MS) in motion scenarios. The IoT device at location \#7 lost connection, so Fig. \ref{fig:distance_performance} displays data sequentially from location \#1 to location \#6.
The results show that distance directly affects signal strength and stability. The received signal strength indicator (RSSI) (top row) decreases as distance increases, with locations like location \#6 experiencing weaker signals and greater instability. Location \#1, being the closest, maintains the strongest RSSI and the most stable connection, underscoring the importance of strategic device placement. Prior work provides theoretical models for how RSSI variance is influenced by human motion and environmental factors~\cite{patwari2011spatial}, aligning with our observations of increased signal fluctuations at farther locations.

CSI amplitude (second row) further highlights this impact. Closer locations exhibit consistent CSI values, while farther locations show greater fluctuations due to signal attenuation and multipath interference. The timestamp difference (third row) reveals increased latency and synchronization issues at greater distances. 
MS (last two rows) demonstrates that closer placements allow for clearer motion detection, as CSI changes are more distinguishable. At greater distances, increased noise makes motion detection more difficult. This aligns with the findings from existing work~\cite{daqing2023placement}, which showed that the optimal transmitter-receiver distance follows a specific pattern, expanding and then contracting the sensing coverage area due to multipath effects. In our case, we observe that locations beyond 10 meters exhibit significant degradation.

The key takeaway is that optimal device placement is critical for reliable CSI-based sensing. Shorter distances ensure stronger signals, better synchronization, and improved motion detection, while increasing distances lead to signal degradation, instability, and reduced sensing accuracy. \emph{
In typical homes, edge IoT devices should be placed within 10 meters of the router to ensure stable sensing quality and sensing performance. The optimal range is less than 6.5 meters, balancing strong RSSI, stable CSI amplitude, and low latency.} Devices should be positioned away from obstacles that cause significant attenuation to ensure good coverage.

\subsection{Sensing Coverage}

\begin{figure}[t]
    \centering
    \begin{subfigure}[b]{0.125\textwidth}
        \centering
      \includegraphics[width=1\textwidth]{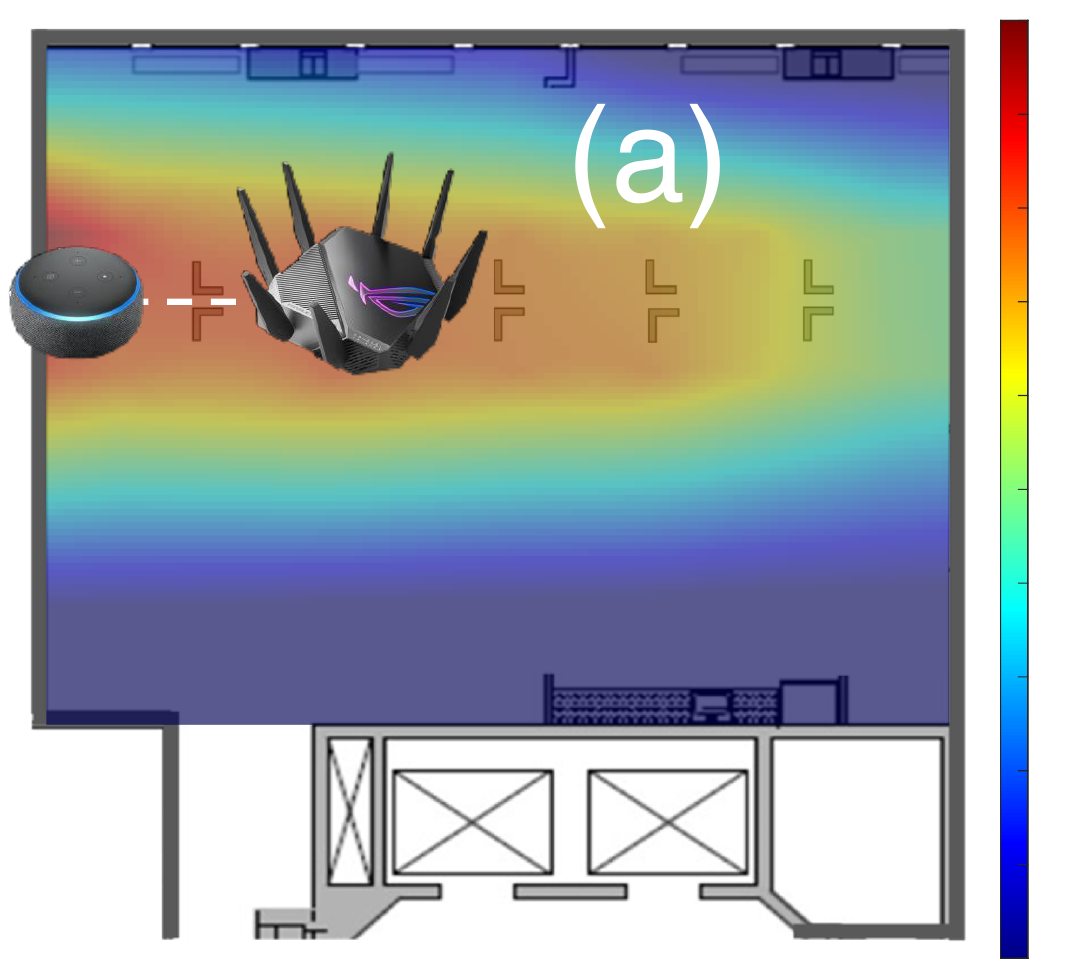}
        \label{fig:coverage_office_single}
    \end{subfigure}%
    \begin{subfigure}[b]{0.18\textwidth}
        \centering
    \includegraphics[width=\textwidth]{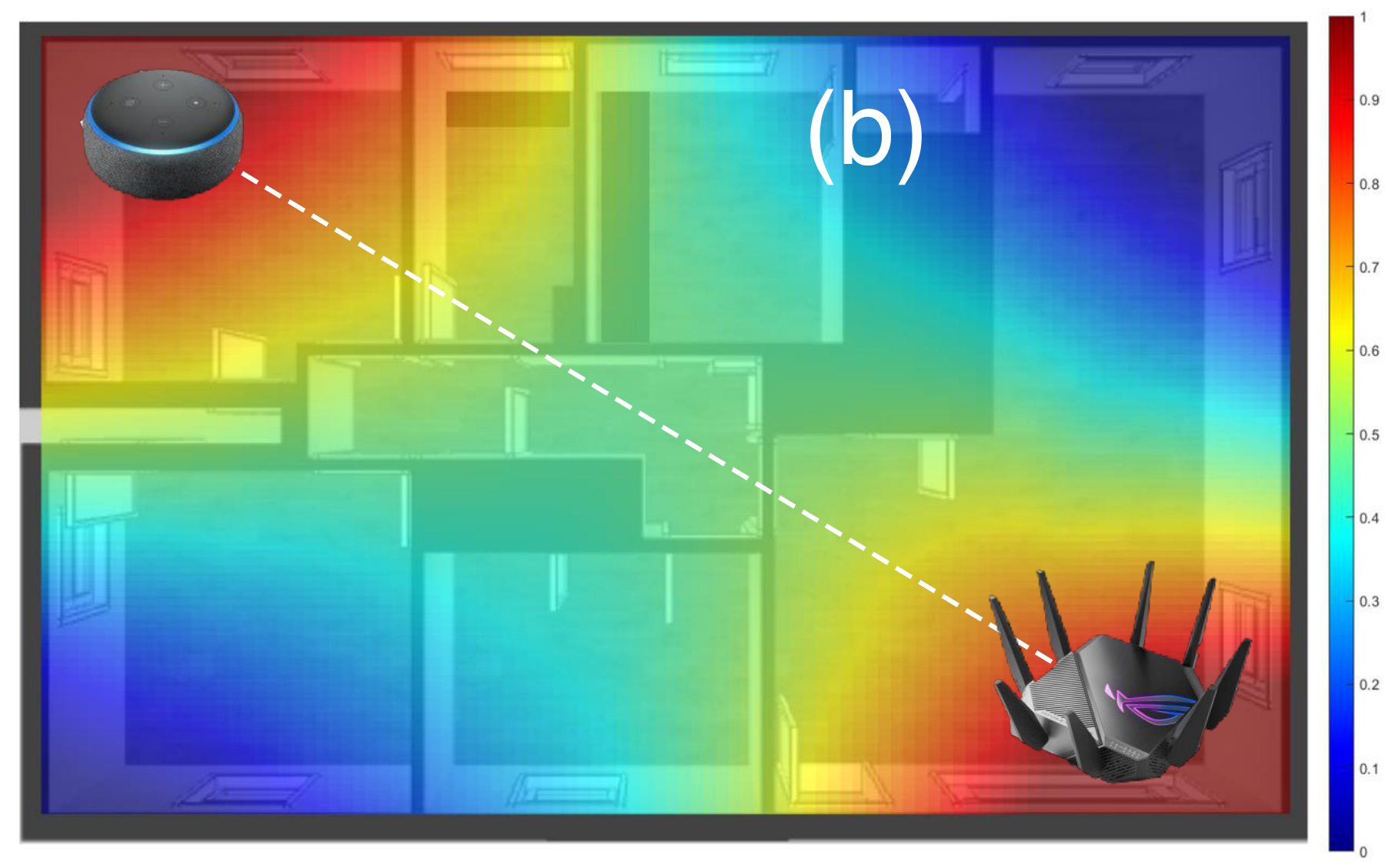}
        \label{fig:coverage_home_single}
    \end{subfigure}%
    \begin{subfigure}[b]{0.182\textwidth}
        \centering
        \includegraphics[width=\textwidth]{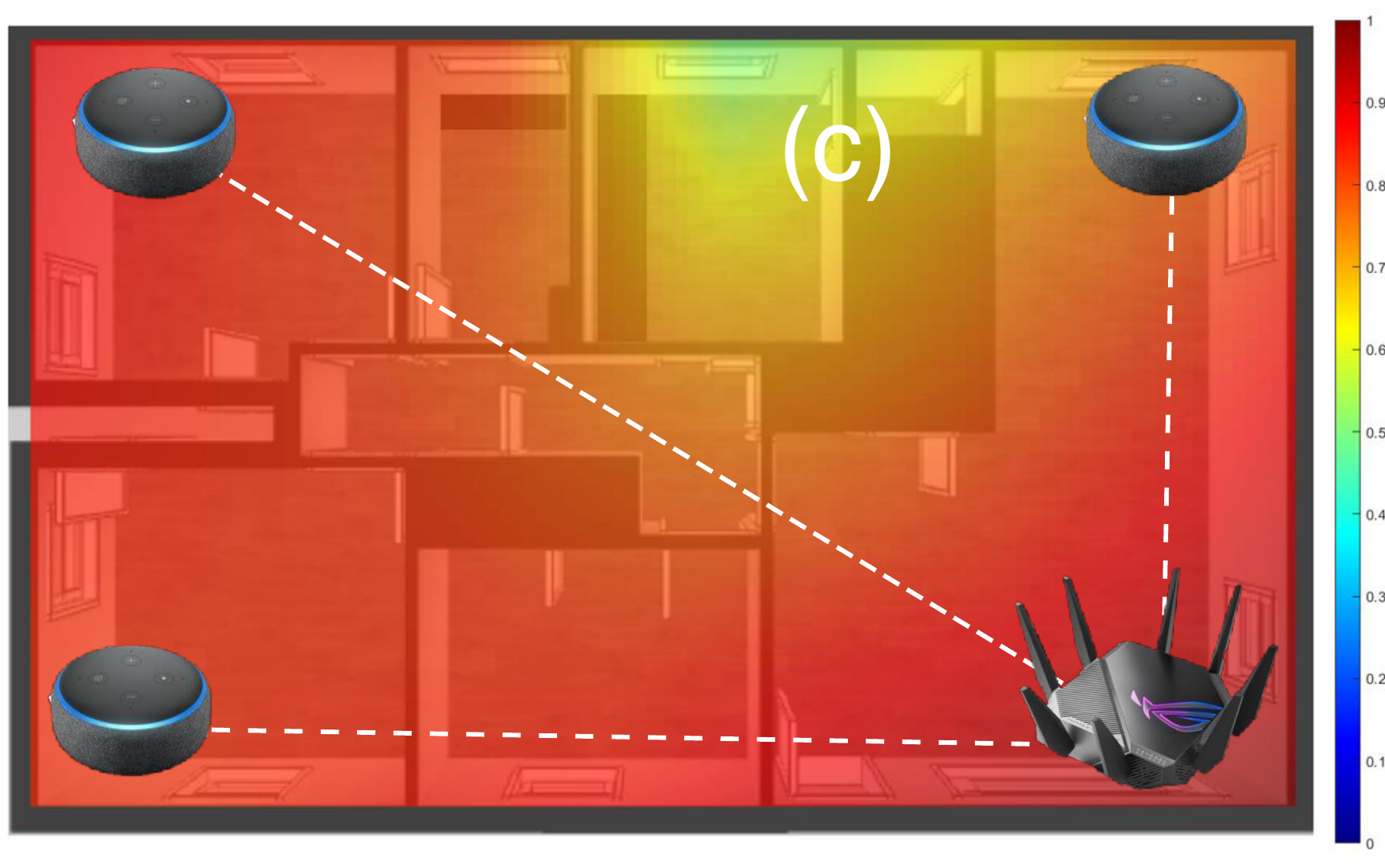}
        \label{fig:coverage_home_multi}
    \end{subfigure}%
    \vspace{-6mm}
    \caption{  Coverage tests conducted in two environments with varying numbers of WiFi links. (a) Single-link coverage in an office environment. (b) Single-link coverage in a home environment. (c) Multi-link coverage in a home environment.}
     
    \label{fig:coverage_test}
\end{figure}
\head{Experiments.} 
Coverage is crucial for a sensing system to track and detect human presence across different regions. We assess coverage through experiments in an office (150 $m^2$) and a home (107 $m^2$), where a subject moves through various areas while the system monitors motion. The router and edge device locations are marked on floorplans (Fig. \ref{fig:coverage_test}).  

To quantify coverage, we define a region's detection probability as the ratio of successful motion detections to the subject's total presence time. A region is considered covered if its detection probability exceeds 80\%.

\head{Findings.}
The results in Fig. \ref{fig:coverage_test}(a) and Fig. \ref{fig:coverage_test}(b) illustrate that with a single pair of WiFi sensing devices, motion near the sensing link can be consistently detected in both office and house environments. However, regions farther from the transmitter-receiver pair experience lower detection rates due to signal attenuation and multipath effects, which weaken sensing capability in obstructed or distant areas. Notably, the measured coverage patterns align well with the theoretical model proposed in previous work~\cite{patwari2011spatial}, where the Cassini oval describes how motion-induced signal strength variance is strongest near the sensing link and diminishes at greater distances.

To improve coverage, we strategically add more IoT sensing devices by leveraging users' existing WiFi devices. As shown in Fig. \ref{fig:coverage_test}(c), \emph{adding one extra device significantly expands the detection area by jointly considering detections from multiple links, resulting in more robust sensing.} This enables more consistent human motion detection, ensuring that as long as one sensing link detects movement, the presence is continuously registered. The combined sensing links merge their detection zones, creating a larger, more reliable coverage area for robust sensing performance.

\section{Multi-user Scenarios}
\label{sec:multi_user}
Multi-user scenarios are critical for real-world WiFi sensing applications, where homes are rarely occupied by a single person. However, they pose significant challenges due to WiFi's low spatial resolution and the complex signal interference caused by multiple moving users. 

\begin{figure*}[t]
    \centering
    \includegraphics[width=1\linewidth]{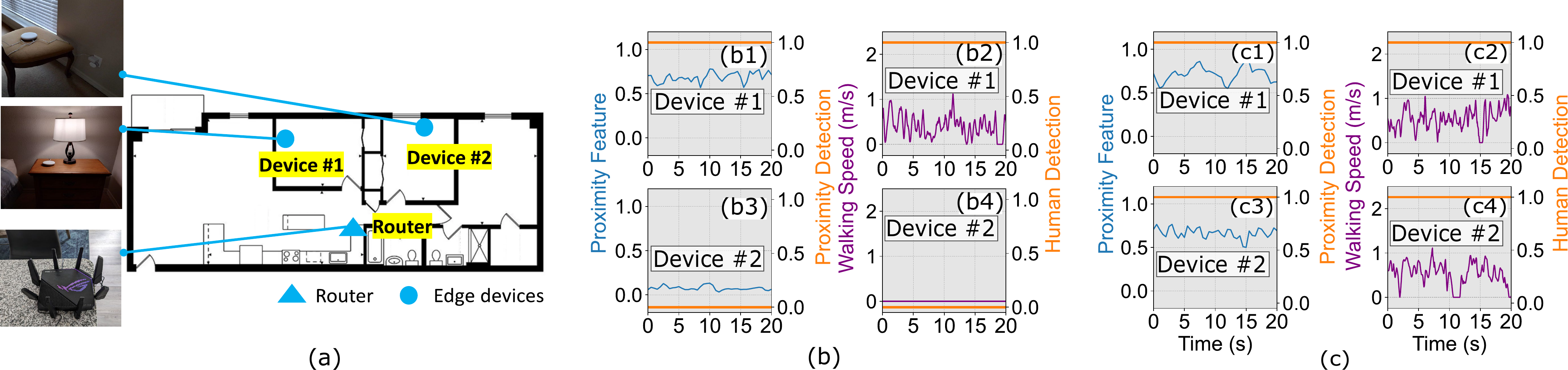}
    \caption{  (a) Device placements. Proximity and speed estimation of Device \#1 and Device \#2 when (b) one user walks around Device \#1, and (c) two users simultaneously walking around Device \#1 and Device \#2, respectively.
     }
      
    \label{fig:multi_user_exp}
\end{figure*}

\head{Challenges.} Multi-user scenarios present a significant challenge in WiFi-based home monitoring due to the complex interactions of multipath propagation and the inherently low spatial resolution of WiFi. Unlike single-user settings, where motion can be directly attributed to a specific occupant, multiple users introduce superimposed signals that are difficult to separate and differentiate. Addressing this challenge requires a refined approach that can effectively identify and track individuals.

An early version of our system aggregated motion signals from all devices without spatial differentiation, assuming total activity could reliably indicate human presence. However, in multi-user homes, this led to frequent misattribution—especially when pets and people moved independently in separate rooms. This limitation motivated a shift to a proximity-based approach that uses relative signal strength and inter-device correlation to associate motion with the nearest device, improving spatial attribution and reducing false detections.

\head{Approaches.} Rather than attempting to resolve the intricate multipath propagation problem directly, we exploit a dense IoT deployment in home environments for motion localization. By analyzing motion patterns in the CSI data and leveraging the spatial distribution of devices, we can effectively distinguish multiple users based on their unique spatial signatures. Recent works~\cite{musefi2023, huang2024wimansbenchmarkdatasetwifibased,person3dyan} demonstrate that near-field WiFi channel variations can be exploited to separate multiple users. Our approach investigates a similar insight as MUSE-Fi~\cite{musefi2023}, yet adopts robust statistical proximity features derived from subcarrier correlation \cite{proximity2023} and gait patterns based on speed estimation\cite{proxmitygait2023}, which have proven effective for determining when an occupant is near a device\cite{proxmitygait2023}. We extend this capability into a distributed IoT framework, enabling occupant-device proximity detection for more accurate multi-user sensing. 
For instance, each device, connected to a central router, monitors its designated area and detects environmental changes when a user enters. By aggregating data from multiple devices, we achieve a more robust multi-user sensing capability. 

\head{Experiments.}
To validate the effectiveness of the proposed method, we conduct an experiment involving two testers in a home setting shown in Fig. \ref{fig:multi_user_exp}(a). The setup includes two edge devices placed in separate rooms, with a central router positioned in the kitchen. Given common practical scenarios where routers are often located in inaccessible positions, we assume individuals primarily move around the edge devices. We consider two scenarios for comparison: \textit{Scenario 1}: A single user moves near a single device; \textit{Scenario 2}: Two users move simultaneously near separate devices.

\head{Findings.}
In Scenario 1, when a user moves near Device \#1, Device \#1 reliably triggers proximity alerts and Device \#1 captures a clear gait pattern from the user's walking speed as shown in Fig. \ref{fig:multi_user_exp}(b1)(b2), while Device \#2 remains inactive.

In Scenario 2, as depicted in Fig. \ref{fig:multi_user_exp}(c1)(c3), proximity detection independently activates for each device based on the local presence of users. After initial detection, gait-based features further enhance human detection, as illustrated by the walking speed analysis in Fig. \ref{fig:multi_user_exp}(c2)(c4). This demonstrates our system's capability to simultaneously and effectively monitor multiple individuals by combining proximity and gait information.

\emph{The integration of proximity and gait features enables robust multi-user localization, mitigating multipath challenges in home environments.} However, our system does not currently identify individual users; when multiple people are near the same device, overlapping motion signals limit differentiation, leaving the single-link multi-user sensing problem an open challenge.

To address this, future work could incorporate identification to improve granularity and enable occupant-specific monitoring. Our system already extracts gait-relevant features such as walking speed and stride periodicity, which prior work~\cite{gaitway} has shown can support identity recognition. Recent deep learning approaches have demonstrated strong performance in user identification by learning discriminative representations of individual movement patterns from raw CSI. The recent large-scale dataset CSI-Bench~\cite{csi-bench} provides rich user identification data across diverse environments, could further support training robust models that generalize across users and homes.
 
Ultimately, combining proximity, gait dynamics, and user identity will be key to achieving fine-grained multi-user tracking in shared spaces.

\section{Deployment on the Edge}
\label{sec:computation}
As the number of connected devices grows and sensing tasks become more complex, the computational burden increases significantly. Efficient deployment is critical for bringing WiFi sensing from lab prototypes to real-world use. Systems must balance real-time responsiveness, privacy, and scalability under tight resource constraints. 

\subsection{Challenges}
WiFi sensing systems must process large volumes of channel measurements to detect and classify motion events. Although on-device (edge) inference can offer immediate response times, many resource-limited edge devices struggle with the computational demands of machine learning models. Meanwhile, offloading computations to the cloud shifts the problem to the network layer, where continuous high-dimensional CSI streaming can saturate residential uplinks. These constraints require an approach that leverages both edge and cloud processing without overburdening either.

\subsection{Computational Constraints}

A straightforward solution is to push the computations to a cloud server. However, based on our extensive user survey, two practical considerations limit this approach: (1) privacy concerns about transferring unprocessed raw data to the cloud, and (2) the need for home monitoring systems-particularly security features like intrusion detection-to function even without internet connectivity. This motivated our move toward hybrid edge–cloud processing. 

\head{Approaches.} To balance the users' concerns and computational limitations, we propose a hybrid edge-cloud approach. Specifically, computationally lightweight modules, including the sensing quality check module, the foundation module, and the proximity module, remain on the edge to ensure motion detection and localization even when offline. More computationally demanding modules are offloaded to the cloud, where only the processed sensing application data is uploaded for advanced tasks like subject recognition when users enable the alarm system.

\head{Experiments.}
To evaluate the feasibility of on-device inference under realistic conditions, we benchmark CPU and memory usage on an Asus GT-AX11000 router (Broadcom 4912 quad-core CPU at 2 GHz, 1 GB RAM). In a star topology with up to 10 IoT devices (Bots) each transmitting CSI packets at 100 Hz, the router served as both data receiver and processor. We implement the SVM-based classifier for motion recognition alongside other essential modules (sensing quality check, foundation, and proximity), and recorded resource usage across 10 repeated trials.

\begin{figure}[t]
    \centering
    \includegraphics[width=1\linewidth]{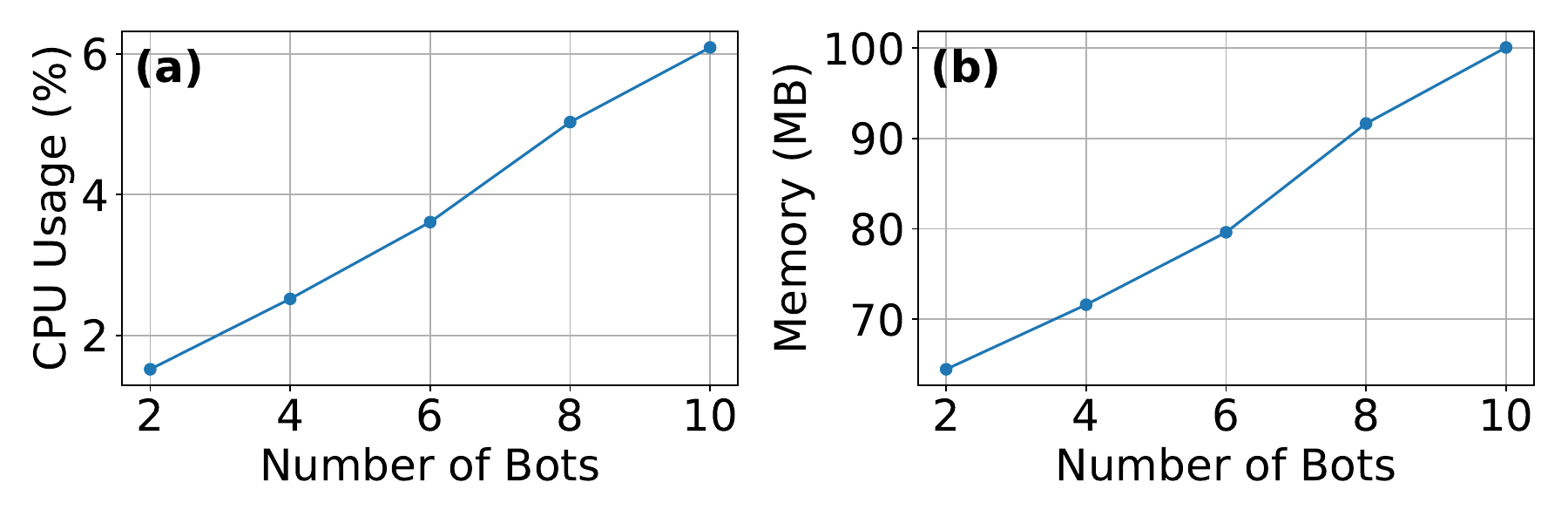}
    \vspace{-7mm}
    \caption{(a) CPU Usage, and (b) Memory usage.}
    \label{fig:edge_constraints}
\end{figure}

Our results show a near-linear increase in CPU and memory usage as the number of Bots increases. Specifically, CPU usage rise from approximately 1.5\% with 2 Bots to 6.09\% with 10 Bots, while memory usage increase from 64 MB to 100 MB, as shown in Fig. \ref{fig:edge_constraints}. While these numbers fall within the hardware capabilities of modern routers, our hardware partner requires that only 7\% of the CPU be allocated to sensing, reserving the rest for networking services. This constraint places a tight upper bound on the local inference capacity.

\head{Findings.}
Current sensing modules can operate within the 7\% CPU usage limit when handling up to 10 IoT devices, but this leaves minimal headroom for supporting additional devices or deploying more computationally intensive models in the future. Deploying more advanced models, such as deep neural networks, is not feasible under this constraint. Furthermore, as IoT devices become more widespread and multi-way sensing develops, real-world deployments involve an increasing number of devices. This leads to higher communication demands and tighter resource budgets per device.

These findings highlight that \emph{even lightweight machine learning models must be carefully selected and optimized for deployment on edge.} The hybrid edge-cloud design strikes a balance: ensure real-time responsiveness and privacy-preserving offline functionality while leveraging the cloud for high-complexity inference tasks. This architecture not only aligns with hardware constraints but also positions the system for future extensibility in diverse sensing applications.

\subsection{Data Transmission Overhead}
The transmission of high-dimensional CSI data from edge devices to the cloud poses a critical scalability challenge in WiFi-based home monitoring systems. 
Although CSI is crucial for motion and activity recognition—capturing fine-grained channel variations across subcarriers, antennas, and time—its volume can overwhelm residential uplinks.

\head{Approaches.}
 To address this bottleneck, we design a hybrid module to reduce data transmission from two key aspects:

\textit{1) Avoiding raw CSI}: Using the established ACF representation from prior work~\cite{widetect,Wispeed2018,wi_moid,SrcSense}, we buffer the ACF queue for a configurable time window and upload only ACF data to the cloud. This approach reduces dimensionality by 61.72\% for one second of 100 Hz CSI (from 66.87 KB to 25.60 KB). Moreover, ACF emphasizes motion-related temporal patterns over static environmental interference and improves cross-environment generalization~\cite{wi_moid,SrcSense}, reducing the need for environment-specific model adaptation.

\textit{2) Event-driven transmission}: To further reduce the data transmission, we only transmit the ACF data when motion is detected. 
By doing so, we considerably reduce data transmission for periods without events of interest.

\head{Experiments.}
Although a single sensing device sampling CSI at 100 Hz with 56 subcarriers and 1 antenna empirically produces only about 149 MB/hour of raw data, in practice, many devices feature more subcarriers (\textit{e.g.}, 200+) and multiple antennas, multiplying the total data rate. Moreover, as the number of IoT sensing devices grows, aggregate traffic can rapidly saturate a typical household's available uplink bandwidth. This heavy load competes with essential activities (\textit{e.g.}, video streaming), causing packet loss, latency spikes, and synchronization errors.

To quantify the effectiveness of our data reduction design, we test the data transmission requirement with 3 users for 21 days in total. Each user has a single access point samples CSI at 100 Hz and 4-10 IoT devices, including both 2.4 GHz and 5 GHz devices, with subcarrier counts ranging from 14 to 196. During the 97 total hours when the users enabled our hybrid module to secure their homes, the average data transmitted per device dropped by 99.72\%—from 1.41 GB/day to just 10.36 MB/day. This substantial decrease alleviates network bottlenecks, especially given the growing prevalence of WiFi-connected IoT devices.

\head{Findings.}
\emph{Compressing the data into ACF form and adopting an event-driven transmission model greatly reduce the data transmission needs.} Moreover, minimizing unnecessary transmissions can lower cloud storage costs and reduce power consumption of edge devices, providing both scalability and sustainability for large-scale WiFi sensing deployments.

\begin{figure}[t]
    \centering
    \includegraphics[width=1\linewidth]{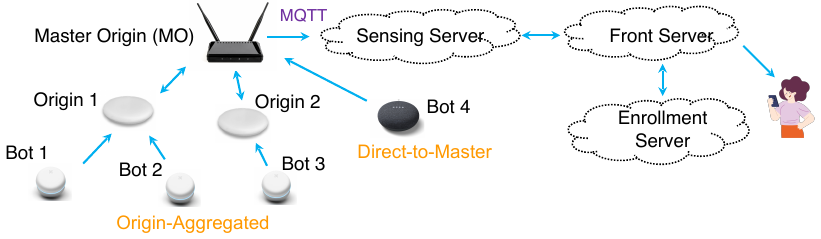}
    \caption{  Overview of deployment architecture.
    }
     
    \label{fig:system_architect}
\end{figure}

\section{Real-World Deployment}
\label{sec:deploy_archit}
Our deployment architecture enables scalable and efficient WiFi sensing across millions of devices. It consists of three core components: on-premise sensing devices, cloud services, and user interfaces (Fig. \ref{fig:system_architect}).

\subsection{Sensing Device Hierarchy}
Central to our design is a concept of flexible device roles, where a single physical device may undertake one or more roles. This flexibility enables deployments to adapt and scale to diverse network topologies and coverage needs.

\head{Bot:}
A WiFi-enabled edge device (\textit{e.g.}, a smart light, speaker) that continuously transmits packets for CSI collection. It can operate on multiple frequency bands (\textit{e.g.}, 2.4 GHz or 5 GHz) as available. Larger deployments may incorporate multiple Bots to enhance sensing coverage and robustness.

\head{Origin:}
A node that receives CSI from one or more Bots, performs sensing health checks, calculates proximity and motion features, and relays these to the Master Origin. In some configurations, an Origin may act as a Bot for another Origin, creating a mesh-like arrangement, and eliminating reliance on dedicated transmitters. A mesh WiFi router or hubs (\textit{e.g.}, Eero, or Google Nest) could serve as an Origin.

\head{Master Origin (MO):}
A centralized hub (\textit{e.g.}, a primary WiFi gateway or router) that fuses feature streams from Origins to infer high-level insights (e.g., presence detection, motion localization). In smaller deployments, the MO may concurrently serve as an Origin.

Devices in our system are organized hierarchically following a Bot–Origin–MO topology, where each Bot performs sensing and communication tasks under the coordination of an Origin. Importantly, an Origin itself can function as a Bot when communicating upward to a higher-tier Origin or the central Master Origin (MO), thereby enabling a flexible, multi-tiered structure that scales across different deployment sizes and layouts.

As illustrated in Fig.~\ref{fig:system_architect}, we identify two common configurations that typically emerge in real-world home environments:
\begin{itemize}
    \item\textbf{Origin-Aggregated Mode:} One or more Bots are connected to a Origin, which aggregates data, performs initial processing, and forwards relevant information upstream to the MO.
    \item \textbf{Direct-to-Master Mode:} Bots communicate directly with the MO without relying on an intermediate Origin.
\end{itemize}


\begin{figure}[t]
    \centering
    \begin{subfigure}[b]{0.25\textwidth}
    \includegraphics[width=\textwidth]{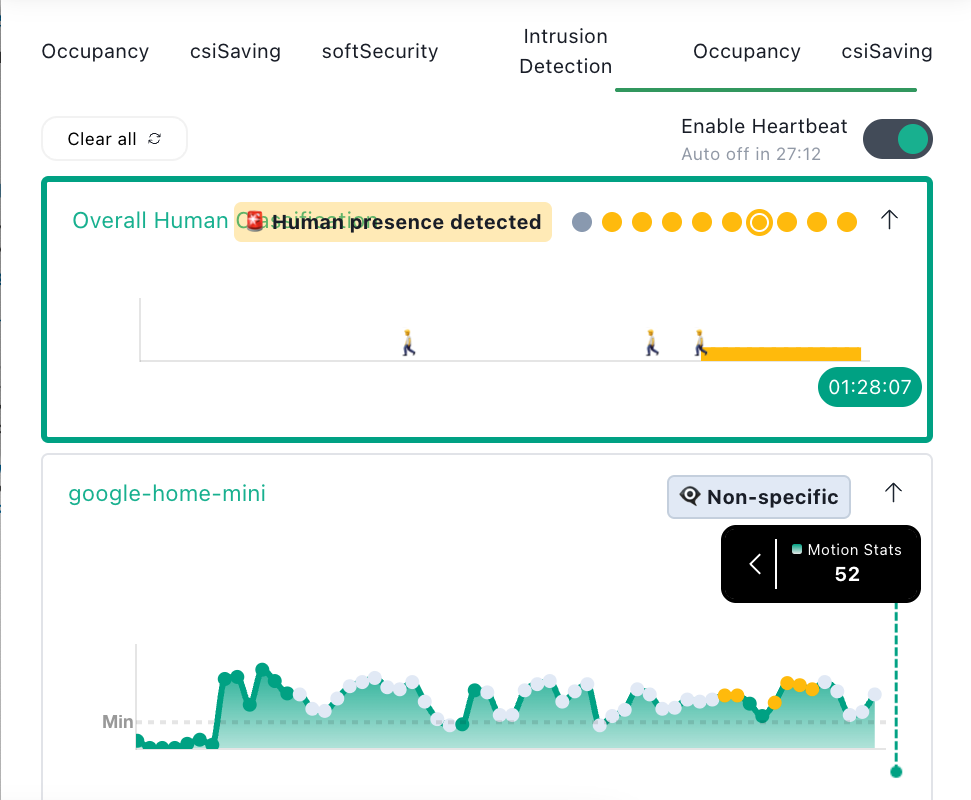}
    \caption{}
    \label{fig:cpu}
  \end{subfigure}
  \hfill
  \begin{subfigure}[b]{0.22\textwidth}
    \includegraphics[width=\textwidth]{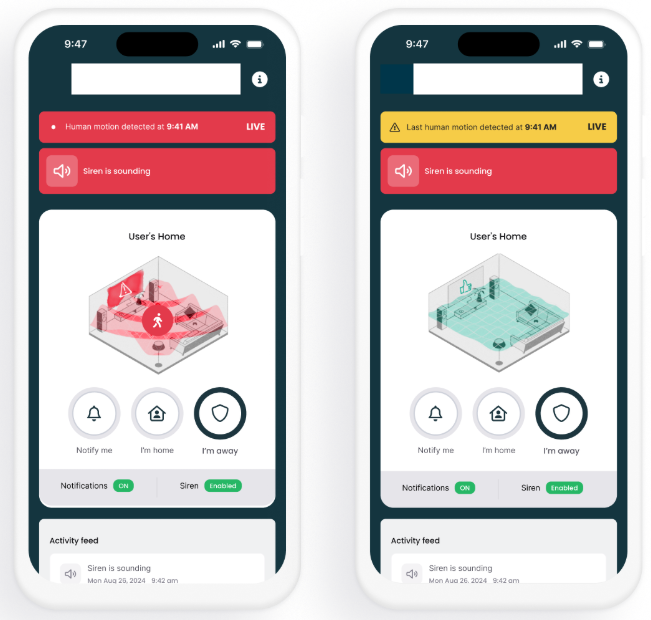}
    \caption{}
    \label{fig:memory}
  \end{subfigure}
    \caption{(a) WebGUI and (b) mobile App.}
    \label{fig:UI}
     
\end{figure}

\subsection{Cloud Infrastructure}
\head{Sensing Server (SS):}
The SS is a cloud-based component dedicated to handling WiFi sensing data flows. It provides message queuing telemetry transport (MQTT)-based connectivity for on-premise devices, relays user-specific information to the Front Server, and maintains minimal historical data for troubleshooting. It remains lightweight on user data to simplify privacy compliance.

\head{Front Server (FS):}
The FS stores user accounts, device configurations, and environment metadata. It retrieves sensing data from the SS, attaches user context, and presents it via user-facing services or analytic tools. It also mediates requests (\textit{e.g.}, configuration changes) between users and the appropriate system components.

\head{Enrollment Server (ES):}
The ES manages device licenses, activation codes, and compliance checks. This layered responsibility ensures robust user/license management with minimal personal data on the SS.

\subsection{User Interface}
The UI enables real-time monitoring, system configuration, and administrative functions. It is implemented as a web-based GUI and a mobile application, illustrated in Fig.~\ref{fig:UI}. By securely communicating with the Front Server over authenticated channels, the interface ensures that users gain immediate visibility into ongoing sensing events while also preserving privacy and access control.

\subsection{Deployment Considerations}
\head{Scalability:} The hierarchical device architecture (Bot → Origin → Master Origin) allows the system to expand in environments with multiple WiFi APs or dense device deployments. Each MO can handle multiple Origins without saturating a single processing node.

\head{Fault Tolerance:} If an MO becomes unavailable, another capable Origin may step in temporarily as the MO, reducing downtime while ensuring quality of service (QoS).

\head{Local vs. Cloud Processing:} Foundation modules run on local devices to reduce backhaul network load. Advanced inference may be local (on the MO) or partially migrated to the cloud if remote computational resources are preferred.

\begin{figure}[t]
    \centering
    \includegraphics[width=1\linewidth]{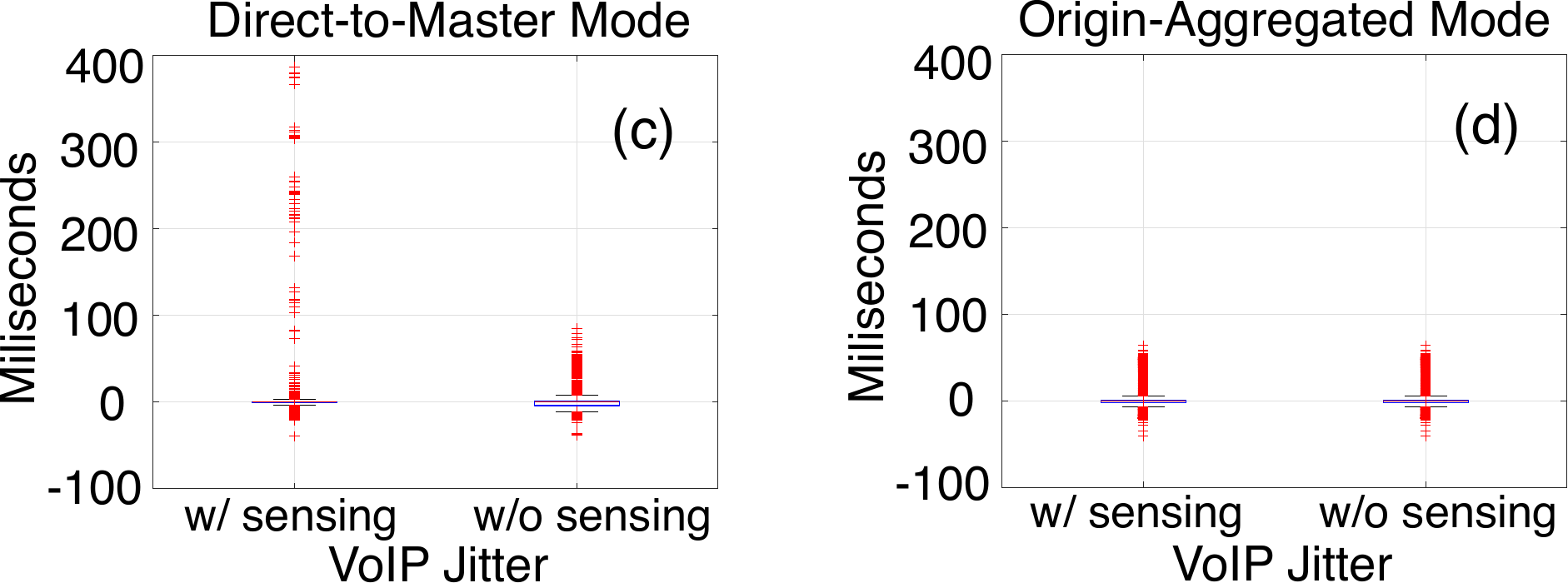}
    \caption{VoIP packet loss and jitter with and without WiFi sensing devices under Direct-to-Master Mode, and Origin-Aggregated Mode.}
         
    \label{fig:VoIPALL}
\end{figure}

\head{Data Protection:} By ensuring that the SS does not store personal user information, the system isolates high-level user data to the FS, simplifying data protection. TLS-based MQTT sessions safeguard transmissions across the internet.

\head{WiFi Sensing Privacy and Security:}
Although WiFi sensing generally raises fewer privacy concerns than video or audio streams, unauthorized access to raw Channel State Information (CSI) can still reveal sensitive information such as vital signs, location patterns, or user activities~\cite{privacyLocation2020}. In terms of security, physical-layer attacks—such as signal jamming, adversarial manipulation of packet preambles~\cite{Li2024}, and covert interference using reconfigurable intelligent surfaces~\cite{RIStealth}—can disrupt sensing performance or mislead machine learning models. Cloud-layer exploits further compound the threat landscape.
To mitigate these risks, recent research has proposed techniques like CSI fuzzing~\cite{Zhang2024Privacy}, which injects artificial noise to obfuscate raw CSI, and adaptive channel hopping~\cite{djuraev2022channel}, which dynamically selects channels to resist jamming. 

Despite growing recognition from standards bodies like IEEE 802.11bf~\cite{IEEE80211bfPrivacy2023}, real-world deployments must still contend with evolving threats, inconsistent standards, and constrained hardware environments. These challenges highlight the need for privacy-preserving mechanisms that are lightweight, adaptable, and compatible with commodity hardware. Coordinated efforts across researchers, manufacturers, and standards organizations are essential to develop scalable, secure WiFi sensing solutions.

\subsection{Implications for Commercial and Industrial IoT Deployments}
While our deployment focuses on residential settings, the architecture and techniques generalize well to commercial and industrial IoT environments such as offices, warehouses, and healthcare facilities.

The Bot–Origin–Master Origin hierarchy supports scalable, fault-tolerant deployments with dense device configurations. Smart lights or access points can serve as Bots, while enterprise routers or gateways act as Origins or Master Origins—enabling modular expansion without saturating any single node.

Our multi-layer sensing quality verification offers a plug-and-play method to assess device suitability across heterogeneous deployments, supporting automated commissioning and robust operation in dynamic environments.

Proximity-based multi-user sensing and biomechanics-informed classification help distinguish humans from common non-human subjects (e.g., pets, robots), enhancing safety and security applications in commercial settings.

Finally, the hybrid edge–cloud architecture and event-driven ACF compression reduce bandwidth and enable offline operation, which is critical in constrained or latency-sensitive environments.
These features make the system broadly applicable to real-world IoT deployments beyond the home.

\section{Joint Sensing and Communication} 
Since our system has been deployed on operational WiFi networks that simultaneously support data communication, we investigate and share our experience on the mutual impacts of WiFi sensing and communication, given the emerging interest in integrated sensing and communication.

\begin{figure}[t]
    \centering
    \includegraphics[width=0.75\linewidth]{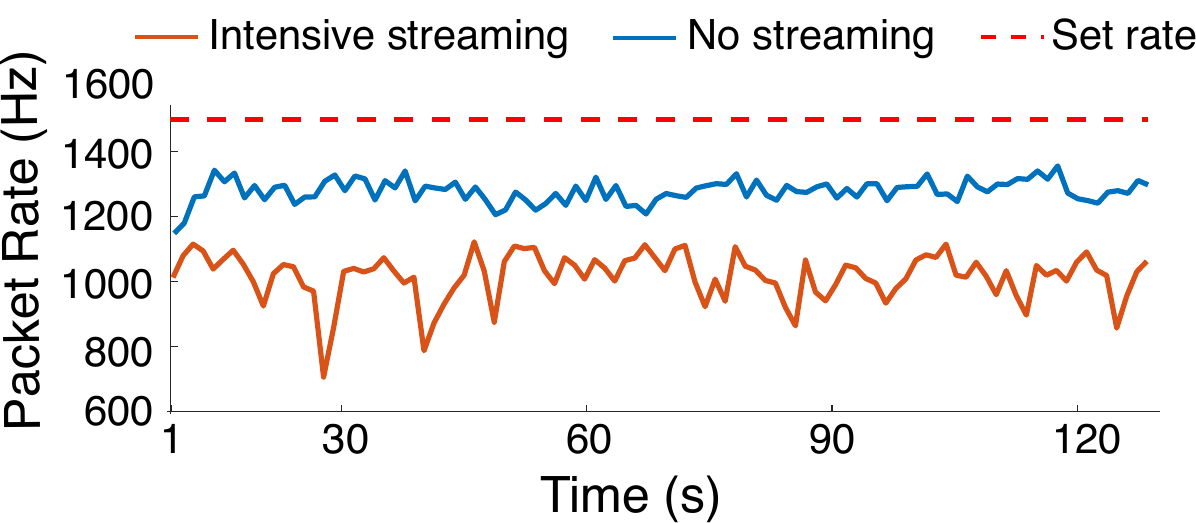}
    \vspace{-4mm}
    \caption{  Impact of streaming on packet loss.}
    \label{fig:soundrate_Impact}

\end{figure}

\label{sec:ISAC}
\head{Challenges.}
WiFi sensing holds great promise, yet may suffer from mutual interference when integrated with data communication in current WiFi standards (802.11n/ac/ax). This issue is especially critical for real-time applications like VoIP and video conferencing, where high jitter degrades user experience. Importantly, the interference is bidirectional: intensive sensing disrupts data communication, while communication traffic can undermine sensing accuracy.

Unlike the emerging IEEE 802.11bf standard~\cite{80211bf}, which dedicates time slots for sensing, current protocols typically rely on periodic null packets to ensure reliable CSI. These dedicated packets for sensing rise Carrier Sense Multiple Access (CSMA)-based collisions, reducing throughput. Concurrently, high-priority traffic (\textit{e.g.}, VoIP) often takes precedence over sensing packets via QoS mechanisms, causing further rate reductions and delays that affect sensing performance.

To quantify the level of interference, we measure VoIP packet jitter under two modes: Direct-to-Master and Origin-Aggregated, using \textit{Flent}~\cite{flent2017} tool.

\head{Impact of WiFi Sensing on Communication.}
Fig. \ref{fig:VoIPALL} shows the effect of high-rate sensing (1500 Hz) on VoIP performance over two minutes with four WiFi sensing devices. In Direct-to-Master mode, these dedicated sensing packets lead to jitter spikes of up to 400 ms, severely degrading VoIP service quality. By contrast, in Origin-Aggregated mode, the Origin consolidates and forwards only processed sensing data to the router, lowering CSI packet influx and mitigating collisions and jitter. This shows that Origin-Aggregated mode can effectively facilitate the integration of WiFi sensing without compromising real-time communication performance.

\head{Impact of WiFi Communication on Sensing.}
We also examine how typical WiFi traffic affects sensing under Direct-to-Master mode. With 4 sensors each transmitting CSI packets at 1500 Hz, the effective packet rate is 1200–1300 Hz under ideal conditions, which is slightly lower than the set rate due to CSMA overhead (Fig. \ref{fig:soundrate_Impact}). Under heavier traffic (\textit{e.g.}, multiple video streams), the packet rate declines by roughly 25\%, to 900–1100 Hz. 
Our results show a trade-off between data traffic for communication and sensing performance, where increased communication traffic compromises sensing bandwidth and reliability.

\head{Findings.}
The mutual interference between sensing and communication over WiFi networks poses a significant challenge for many prior approaches that require high CSI sampling rates \cite{high_rate1,high_rate2,XModal_ID_Mostofi, widar2,pose_highrate}. 
Nevertheless, \emph{our deployment uses a low CSI rate of 100 Hz, allowing it to run smoothly on home WiFi networks and scale over millions of devices as the interference is negligible.} 
Our results also show that, under legacy WiFi protocols, adopting flexible deployment strategies can balance sensing accuracy and data communication. Specifically, the Direct-to-Master mode is well-suited for low-traffic environments, ensuring high packet rates and simpler infrastructure, while the Origin-Aggregated mode excels under heavy traffic, centralizing data processing to reduce interference and maintain network efficiency. 
Future WiFi standards featuring integrated sensing and communication design will address the issue further.

\section{Conclusion}
\label{sec:con}
This paper shares our experience and findings learned from the world's first multi-year, millions-scale WiFi-based home monitoring systems in real-world settings. We identify and investigate critical challenges in motion sources, hardware heterogeneity, multi-user sensing, edge deployment, and joint sensing and communication. 
To the best of our knowledge, our system is the first WiFi sensing solution that has been deployed at scale for real-world applications.
We hope our experience can encourage future efforts to develop practical WiFi sensing systems that bridge the gap between laboratory concepts and real-world applications, ultimately enhancing everyday life through advanced connectivity and sensing intelligence. 

\section*{References}
\bibliographystyle{IEEEtran}
\bibliography{reference,reference-gz}

\vfill

\end{document}